\documentclass[11pt,a4paper]{article}
\pdfoutput=1

\usepackage{jheppub}

\usepackage{amsfonts}
\usepackage{mathtools}
\usepackage{amssymb}
\usepackage{eucal}
\usepackage{mathrsfs}
\usepackage{slashed}
\usepackage{amsmath}
\usepackage{braket}
\usepackage{graphicx}
\usepackage{float}
\usepackage{subfloat}
\usepackage{relsize}
\usepackage[font=scriptsize,labelfont=bf]{caption}
\usepackage{subcaption}
\usepackage{xfrac}
\usepackage{enumerate,mdwlist}
\usepackage{ifthen}
\usepackage[dvipsnames]{xcolor}
\usepackage{tikz}
\usepackage{xspace}

\DeclareMathOperator{\keV}{keV}

\DeclareMathOperator{\GeV}{GeV}
\DeclareMathOperator{\TeV}{TeV}

\DeclareMathOperator{\LH}{L}

\DeclareMathOperator{\PRH}{P_R}
\DeclareMathOperator{\PLH}{P_L}

\newcommand{\ie}{{\em i.e.}\xspace}
\newcommand{\eg}{{\em e.g.}\xspace}

\newcommand{\rhs}{RHS\xspace}

\newcommand{\lrb}[1]{\left( #1 \right)}

\newcounter{NumArgs}

\newcommand{\eqs}[1]{\setcounter{NumArgs}{0}\foreach\i in{#1}{\stepcounter{NumArgs}}%
	\ifthenelse{\equal{\theNumArgs}{1}}{eq.~(\ref{#1})}%
	{\ifthenelse{\equal{\theNumArgs}{2}}%
		{eqs.~\foreach\i[count=\q]in{#1}{\ifthenelse{\equal{\q}{\theNumArgs}}{and (\ref{\i})}{(\ref{\i})~}}}%
		{eqs.~\foreach\i[count=\q]in{#1}{\ifthenelse{\equal{\q}{\theNumArgs}}{and (\ref{\i})}{(\ref{\i}),~}}}}}

\newcommand{\Eqs}[1]{\setcounter{NumArgs}{0}\foreach\i in{#1}{\stepcounter{NumArgs}}%
	\ifthenelse{\equal{\theNumArgs}{1}}{Eq.~(\ref{#1})}%
	{\ifthenelse{\equal{\theNumArgs}{2}}%
		{Eqs.~\foreach\i[count=\q]in{#1}{\ifthenelse{\equal{\q}{\theNumArgs}}{and (\ref{\i})}{(\ref{\i})~}}}%
		{Eqs.~\foreach\i[count=\q]in{#1}{\ifthenelse{\equal{\q}{\theNumArgs}}{and (\ref{\i})}{(\ref{\i}),~}}}}}

\newcommand{\refs}[1]{\setcounter{NumArgs}{0}\foreach\i in{#1}{\stepcounter{NumArgs}}%
	\ifthenelse{\equal{\theNumArgs}{1}}{(\ref{#1})}%
	{\ifthenelse{\equal{\theNumArgs}{2}}%
		{\foreach\i[count=\q]in{#1}{\ifthenelse{\equal{\q}{\theNumArgs}}{and (\ref{\i})}{(\ref{\i})~}}}%
		{\foreach\i[count=\q]in{#1}{\ifthenelse{\equal{\q}{\theNumArgs}}{and (\ref{\i})}{(\ref{\i}),~}}}}}

\newcommand{\Figs}[1]{\setcounter{NumArgs}{0}\foreach\i in{#1}{\stepcounter{NumArgs}}%
	\ifthenelse{\equal{\theNumArgs}{1}}{Fig.~(\ref{#1})}%
	{\ifthenelse{\equal{\theNumArgs}{2}}%
		{Figs.~\foreach\i[count=\q]in{#1}{\ifthenelse{\equal{\q}{\theNumArgs}}{and (\ref{\i})}{(\ref{\i})~}}}%
		{Figs.~\foreach\i[count=\q]in{#1}{\ifthenelse{\equal{\q}{\theNumArgs}}{and (\ref{\i})}{(\ref{\i}),~}}}}}

\newcommand{\Gen}[2]{\setcounter{NumArgs}{0}\foreach\i in{#2}{\stepcounter{NumArgs}}%
	\ifthenelse{\equal{\theNumArgs}{1}}{#1.~(\ref{#2})}%
	{\ifthenelse{\equal{\theNumArgs}{2}}%
		{#1.~\foreach\i[count=\q]in{#2}{\ifthenelse{\equal{\q}{\theNumArgs}}{and (\ref{\i})}{(\ref{\i})~}}}%
		{#1.~\foreach\i[count=\q]in{#2}{\ifthenelse{\equal{\q}{\theNumArgs}}{and (\ref{\i})}{(\ref{\i}),~}}}}}

\title{Ultraviolet freeze-in baryogenesis}

\author[1]{Andreas Goudelis}
\author[2]{\!\!, Dimitrios Karamitros}
\author[3]{\!\!, Pantelis Papachristou}
\author[3]{\!\!, Vassilis C. Spanos}
\affiliation[1]{Laboratoire de Physique de Clermont (UMR 6533), CNRS/IN2P3, Univ.\ Clermont Auvergne, 4 Av.\ Blaise Pascal, F-63178 Aubi\`ere Cedex, France}
\affiliation[2]{School of Physics and Astronomy, The University of Manchester, Manchester M13 9PL, United Kingdom}
\affiliation[3]{National and Kapodistrian University of Athens, Department of Physics, Section of Nuclear and  Particle Physics, GR-157 84 Athens, Greece}
\emailAdd{andreas.goudelis@clermont.in2p3.fr}
\emailAdd{dimitrios.karamitros@manchester.ac.uk}
\emailAdd{pantelisp@phys.uoa.gr}
\emailAdd{vspanos@phys.uoa.gr}

\abstract{We study a mechanism through which the cosmic dark matter density can be explained simultaneously with the observed baryon asymmetry of the Universe. At the core of our proposal lie the out-of-equilibrium scattering processes of bath particles which are responsible for the production of feebly-interacting dark matter. The same processes violate $CP$, which further leads to an asymmetry between matter and antimatter being generated in the visible sector. We focus on the possibility that these interactions are described through non-renormalizable operators, which leads to both dark matter and the baryon asymmetry being produced at high temperatures. The mechanism is exemplified by studying two concrete scenarios, one involving scalar and one involving fermion dark matter. We find that in both cases it is, indeed, possible to achieve a common explanation for the dark matter content and the matter-antimatter asymmetry of the Universe, provided that dark matter is in the keV mass range.}

\begin{document}
\maketitle

\section{Introduction}\label{sec:intro}

The nature of dark matter (DM) \cite{Mambrini2021} and the origin of the matter-antimatter asymmetry of the Universe constitute two of the most fundamental open questions in contemporary high-energy physics and cosmology. Despite the fact that, to date, there is no compelling argument as to why (and how) these two questions could be connected with each other, the idea that they might admit a common microscopical and cosmological explanation has been consistently entertained in several different frameworks. In the context of asymmetric DM scenarios (for reviews \textit{cf e.g.} \cite{Davoudiasl_2012,Petraki:2013wwa}), the observed DM cosmic abundance is viewed as being due to an asymmetry being created between the abundances of DM and dark anti-matter which survived after the symmetric component annihilated away, in a manner similar to (and, typically, connected with) the asymmetry between baryons and anti-baryons in the visible sector. In scenarios of baryogenesis via weakly interacting massive particles (``WIMPy baryogenesis'') \cite{McDonald_2011,Cui_2012,Cui_2013,Chu:2021qwk}, on the other hand, there is no asymmetry in the dark sector and the observed baryon asymmetry is generated during the freeze-out of DM particles, once they fall out of equilibrium with the thermal plasma.

In \cite{Hall:2010jx} it was pointed out that in DM physics there exists (at least) one framework in which the out-of-equilibrium Sakharov condition \cite{Sakharov:1967dj} is, in a sense, built-in from the start: the freeze-in mechanism \cite{McDonald:2001vt,Hall:2009bx}. In this context, which was further critically elaborated upon, \textit{e.g.}, in \cite{Hook:2011tk,Unwin:2014poa}, the authors suggested that viable asymmetries could be generated (frozen-in) simultaneously in the dark and visible sectors through out-of-equilibrium decays or annihilations of bath particles. The framework outlined in \cite{Hall:2010jx} involves asymmetric DM, \textit{i.e.} generating asymmetries both in the dark and in the visible sector through a common mechanism. 

Instead, in \cite{Shuve_2020} another possibility was investigated: that of freezing-in a symmetric DM abundance, along with a baryon asymmetry in the visible sector. In its original version (\textit{cf} also \cite{Berman:2022oht}) the baryon asymmetry generation is achieved through DM oscillations in a manner similar to the situation occurring in the Akhmedov-Rubakov-Smirnov (ARS) leptogenesis \cite{Akhmedov:1998qx}. In \cite{Goudelis:2021qla} a different approach was proposed, according to which DM and a baryon asymmetry are simultaneously generated through $CP$-violating decays of heavy bath particles, with $CP$ violation arising through the interference between tree-level and one-loop diagrams in a manner more similar to Dirac leptogenesis \cite{Dick:1999je,Murayama:2002je,Cerdeno:2006ha,Gonzalez-Garcia:2009cza}. A similar idea was also recently investigated in \cite{Chand:2022vrf}.

A common feature of these proposals is that they operate at relatively low temperatures. Indeed, the models proposed in Refs.~\cite{Shuve_2020,Berman:2022oht,Goudelis:2021qla,Chand:2022vrf} are renormalizable, which leads to the freeze-in production (both of DM and of the baryon asymmetry) being infrared (IR)-dominated. However, it is known that in theories involving non-renormalizable interactions parametrized by higher-dimensional Lagrangian operators, freeze-in production occurs predominantly in the ultraviolet (UV) regime, \textit{i.e.} close to the highest considered temperature. In the context of DM, this ``ultraviolet freeze-in'' scenario was first analyzed in \cite{Elahi:2014fsa}. On the side of baryogenesis, relevant constructions were presented in \cite{Baldes:2014rda,Baldes:2014gca} (and UV-completed in \cite{Baldes:2015lka}), in which the authors were mainly interested in studying scenarios in which the baryon asymmetry of the Universe is generated through scattering processes.

In this paper we study whether a viable (symmetric) DM abundance and a matter-antimatter asymmetry can be simultaneously frozen-in in the UV through non-renormalizable interactions. Our study will focus on two cases, those of a scalar and a fermionic DM candidate, which will lead us to consider operators of different dimensionality, amounting to quantitatively (albeit not qualitatively) different results. We will see that in such a framework it is, indeed, possible to explain both observations simultaneously as long as DM is relatively light, while being consistent with different cosmological constraints.

The paper is structured as follows: In Section \ref{sec:general} we recall some key elements concerning the freeze-in DM production mechanism (notably its UV-dominated version) and we discuss different constraints on the DM properties that will be of relevance for the following. In Section \ref{sec:models} we study two concrete realizations of our ``ultraviolet freeze-in baryogenesis'' mechanism, one in which DM will be assumed to be a scalar and one in which it will be taken to be a Dirac fermion. Lastly, Section \ref{sec:conclusions} contains our conclusions, whereas the Appendix that follows contains some more technical aspects of our analysis.

\section{Ultraviolet freeze-in and constraints on the DM mass}\label{sec:general}

Before studying concrete models, let us briefly recall some basic features of frozen-in DM and discuss a set of phenomenological constraints which will be important for the analysis that follows. Some additional issues which are more model-dependent will be discussed later on, at the appropriate sections.

\subsection{Freeze-in DM production through non-renormalizable operators}

The freeze-in DM production mechanism \cite{McDonald:2001vt,Hall:2009bx} relies on two basic features:
\begin{itemize}
 \item The initial DM abundance is negligible.
 \item DM only interacts extremely weakly (``feebly") with all particles in the thermal bath.
\end{itemize}
These two assumptions in conjunction allow us to consider only DM production processes and to ignore backreactions from the DM sector towards the visible one, \textit{i.e.} DM annihilation processes. 

Moreover, as pointed out \textit{e.g.} in \cite{Hall:2009bx}, if the two sectors only interact via renormalizable operators then DM production is a mostly low-temperature (``infrared'') effect. As an example, if DM is produced through the annihilation of light bath particles, its production peaks around a temperature $T \sim m_{\text{DM}}/3 $. 

In this work we consider, instead, another possibility, in which an effective operator $\mathcal{\hat{O}}_{\left(n\right)}$ of dimension $(n+4)$
\begin{equation}
\mathcal{L}\,\supset\,\frac{1}{\Lambda^n}\,\mathcal{\hat{O}}_{\left(n\right)} \;,
\end{equation}
where $n=1,2,\ldots$ and $\Lambda$ is the energy scale of the effective field theory (EFT), is responsible for the generation of the freeze-in DM abundance (and, as we will see, also for the baryon asymmetry of the Universe). In this case DM is mostly produced in the ultraviolet regime, \textit{i.e.} close to the reheating temperature $T_{\rm RH}$. 

Following the analysis of~\cite{Hall:2009bx,Elahi:2014fsa}, and assuming that DM is produced through the annihilation of highly relativistic particles, the matrix element squared of the corresponding DM production process is expected to scale as
\begin{equation}
    \sum_{\rm idof}|\mathcal{M}|^2\, = A\, \lrb{\frac{\tilde{s}}{\Lambda^2}}^{n} \;,
\label{eq:non_renorm_M_DM}
\end{equation}
where $\sqrt{\tilde{s}}$ is energy in the center-of-mass frame and the sum runs over all internal degrees of freedom (without averaging), $n=d-4$, and $A$ is a model-dependent constant.   

In the types of models that we will study in this paper, DM ($\chi$) and its antiparticle ($\bar \chi$) are produced via pair-annihilations of particles (say, $b_1$ and $b_2$) in thermal equilibrium with the plasma; \ie via processes of the type $b_1 \, b_2 \to \chi \, \chi$ and $\bar b_1 \, \bar b_2 \to \bar \chi \, \bar \chi$. In this case, the Boltzmann equation that describes the evolution of the number of DM particles takes, in the limit that $\Lambda$ is larger than all the masses $m_i$ involved in the process, the form \cite{Hall:2009bx,Elahi:2014fsa}
\begin{align}
    \dfrac{d Y_{\rm DM}}{d T} \approx  - 2 A \, \frac{4^{n+1} n! (n+1)!}{ 1024\times 1.66\, \pi^7 } \dfrac{45}{g_{\star s}\sqrt{g_{\star \rho}}}\frac{M_{Pl} \,  T^{2n-2}}{\Lambda^{2n}} \; ,
\label{eq:DM_Boltzmann}
\end{align}
where $Y_{\rm DM}  \equiv \lrb{n_{\chi} + n_{\bar \chi}}/s$ is the DM yield (number density $n_\chi$ over the entropy density $s=2\pi^2g_{\star s}T^3/45$), $M_{Pl} \approx 1.22 \times 10^{19}~\GeV$ is the Planck mass and $g_{\star s},\,g_{\star\rho}$ are the relativistic degrees of freedom of the plasma with respect to the entropy and energy densities -- assumed to be constant at the temperatures of interest, respectively.\footnote{The factor $2$ is included because both $\chi$ and $\bar \chi$ are assumed to be produced in equal amounts.}

The form of \Eqs{eq:DM_Boltzmann} implies that DM production becomes maximal when the plasma attains its highest temperature. Integrating, then, \Eqs{eq:DM_Boltzmann} from the reheating temperature\footnote{Note that in some frameworks DM production may also peak at temperatures $T > T_{\rm RH}$, \textit{i.e.} during the reheating period \cite{Garcia:2017tuj}. This happens, typically, when considering effective operators of higher dimension than the ones that we will consider in this paper.} $T_{\rm RH}$ down to some temperature $T<T_{\rm RH}$, the DM yield becomes 
\begin{align}
    Y_{\rm DM}(T) \approx  2 A \, \frac{4^{n+1} n! (n+1)!}{ 2n-1 } \dfrac{45}{ 1024\times 1.66 \, \pi^7 g_{\star s}\sqrt{g_{\star\rho}}}
    \frac{M_{Pl} \,  (T_{\rm RH}^{2n-1} -  T^{2n-1})}{\Lambda^{2n}} \; ,
\label{eq:YDMT}
\end{align}
which leads to the DM relic abundance\footnote{Here we ignore a term proportional to the present-day temperature, since the latter is negligible compared to $T_{\rm RH}$.}
\begin{align}
    \Omega_{\rm DM} h^2 \approx  \dfrac{2.4 \times 10^{23}}{g_{\star s}\sqrt{g_{\star\rho}}} \  
    \dfrac{A \, 4^n  \,n!  \, (n+1)!}{2n-1}  \,  
    \dfrac{T_{\rm RH}^{2n-1} \ m_{\rm DM}}{\Lambda^{2n}}  \; .
\label{eq:relic_general}
\end{align}
From \Eqs{eq:relic_general} we can see that we, indeed, recover the well-known result; namely if DM is produced through the action of non-renormalizable operators its predicted freeze-in abundance scales as a power of the reheating temperature, with the exponent depending on the operator's effective dimension. In Section \ref{sec:models} we will make extensive use of this expression, specializing it to different types of higher-dimensional operators.

\subsection{Constraints on the DM mass}

As we will see in the following section, requiring the simultaneous generation of the observed DM abundance in the Universe along with a viable baryon asymmetry will lead us to consider fairly low DM masses. Before proceeding further with our analytical and numerical treatment let us, hence, briefly discuss two relevant constraints on the DM mass.

\subsubsection{Non-equilibration of the visible and dark sectors}

The main premise of freeze-in is that the visible and dark sectors remain out-of-equilibrium with each other during the cosmic expansion. In the ultraviolet freeze-in case the condition of non-equilibration between the two sectors can be formulated by requiring that the abundance of the DM at $T_{\text{RH}}$ be much smaller than its equilibrium value \cite{Elahi:2014fsa}, \textit{i.e.}
\begin{equation}\label{eq:out-of-equilibrium constraint}
Y_{\text{DM}}\left(T_{\text{RH}}\right)\,\ll\,Y_{\text{DM}}^{\text{eq}}\left(T_{\text{RH}}\right) \; .
\end{equation}
As $T_{\text{RH}}\gg m_{\text{DM}}$ the DM can be treated as effectively massless and its equilibrium abundance at $T_{\text{RH}}$ can be approximated by $Y_{\text{DM}}^{\text{eq}}(T_{\text{RH}})\approx \zeta(3)T^3_{\text{RH}}/(\pi^2 s(T_{\text{RH}}))$. Since the DM abundance effectively freezes in close to $T_{\text{RH}}$ and, subsequently, remains quasi-constant, it holds that $Y_{\text{DM}}\left(T_{\text{RH}}\right)\approx Y_{\text{DM}}\left(T_{0}\right)$, where $T_0\approx 2.75 \text{K}$ is the temperature at the present day. The DM relic density is, in turn, given by $\Omega_{\text{DM}}=m_{\text{DM}}s_0Y_{\text{DM}}(T_{0})/\rho_c$, where $\Omega_{\text{DM}}h^2= 0.1200\pm 0.0012$, $\rho_c\equiv 3H_0^2/8\pi\,G\approx 10.537\,h^2\,\GeV m^{-3}$ is the critical density and $s_0\approx 2.9\times 10^{9}\,m^{-3}$ is the entropy density at the present day \cite{Planck:2018vyg}. Hence, the out-of-equilibrium constraint of Eq.~\eqref{eq:out-of-equilibrium constraint} imposes a lower bound on the DM mass
\begin{equation}
m_{\text{DM}}\,\gg\,\frac{\Omega_{\text{DM}}\,\rho_c}{Y_{\text{DM}}^{\text{eq}}\left(T_{\text{RH}}\right)\,s_0}\,\approx\,0.17\,\keV
\end{equation}
where we have used that in the Standard Model (SM) $g_{\star s}\left(T_{\text{RH}}\right)=106.75$.

\subsubsection{Lyman-$\alpha$ constraint}

\begin{figure}[t]
\centering
\includegraphics[width=0.7\linewidth]{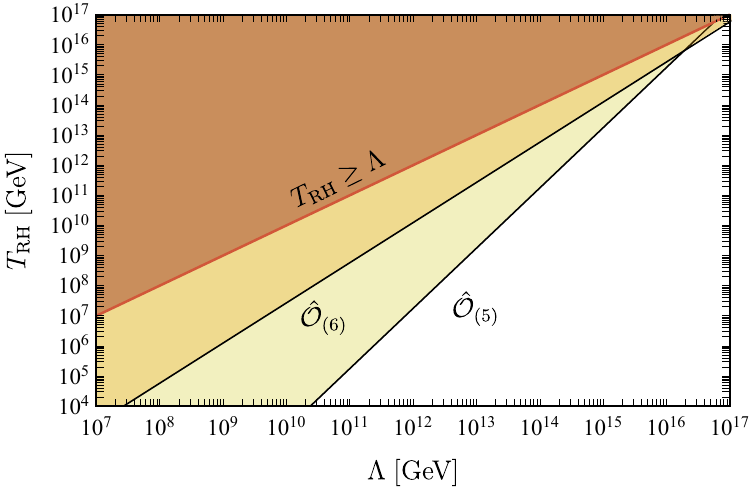}
\caption{The Lyman-$\alpha$ constraint (light and dark yellow-shaded regions) mapped onto the $\left(\Lambda,\,T_{\text{RH}}\right)$ plane for the $\mathcal{\hat{O}}_{\left(5\right)}$ and $\mathcal{\hat{O}}_{\left(6\right)}$ operators. Each line corresponds to the observed DM relic density for the lowest allowed value of the DM mass, as given by Eqs.~\eqref{eq:Lyman-a dim5} and \eqref{eq:Lyman-a dim6}, respectively. The light brown-shaded region is inconsistent with our assumptions, as it requires knowledge of the full UV-completed theory.}
\label{fig:constraints}
\end{figure}

A more stringent constraint on the DM mass comes from Lyman-$\alpha$ forest observations. For frozen-out DM candidates the current bound reads $m_{\text{DM}}\gtrsim (1.9-5.3)\,\keV$ at $95\%$ C.L. \cite{Garzilli:2019qki,Palanque-Delabrouille:2019iyz,Irsic:2017ixq}. This limit has been mapped onto different cases of freeze-in-produced DM, depending on the type of the relevant interactions, on the dimension of the (effective) operator involved and on the statistical distribution of the various states. In the case of $2\leftrightarrow 2$ scatterings with two fermions annihilating into two DM states via dimension-5 ($\mathcal{\hat{O}}_{\left(5\right)}$) and dimension-6 ($\mathcal{\hat{O}}_{\left(6\right)}$) operators, the Lyman-$\alpha$ lower bound on the DM mass has been shown to translate, respectively, to \cite{Ballesteros:2020adh}:
\begin{subequations}
\begin{align}
\mathcal{\hat{O}}_{\left(5\right)}:&\quad m_{\text{DM}}\,\gtrsim\,
    \left(4.0 - 15.5\right)\,\keV\label{eq:Lyman-a dim5}
\\\nonumber\\
\mathcal{\hat{O}}_{\left(6\right)}:&\quad m_{\text{DM}}\,\gtrsim\,
    \left(4.6 - 18.1\right)\,\keV \; . \label{eq:Lyman-a dim6}
\end{align}
\end{subequations}
The impact of these bounds is shown in Fig.~\ref{fig:constraints}, in which we depict the $(\Lambda, T_{\rm RH})$ combinations for which it is possible to explain the observed DM abundance in the Universe, while simultaneously satisfying the conditions of Eqs.~\eqref{eq:Lyman-a dim5} and \eqref{eq:Lyman-a dim6}.

\section{Concrete Realizations of UV freeze-in baryogenesis}\label{sec:models}

Let us now examine, as a proof-of-concept, two concrete examples in which the observed DM abundance and the matter-antimatter asymmetry of the Universe can be simultaneously explained through the action of non-renormalizable operators.

\subsection{Scalar DM}\label{sec:scalarDM}

We consider a simple extension of the SM by two heavy vector-like leptons $F_{1,2}$, which are singlets under $SU(3)_c\times SU(2)_{\LH}$ but carry hypercharge, and by a complex gauge-singlet scalar $\varphi$ which will play the role of our frozen-in DM candidate. We will focus on the following interaction Lagrangian~\footnote{Note that the symmetries allow additional operators, including the -- lower dimension -- Higgs portal~\cite{Burgess:2000yq} (for a review on its implications see, \eg~\cite{Lebedev:2021xey}). A study including such operators is beyond the scope of this work.}
\begin{equation}\label{eq:Lagdim5}
\mathcal{L}_{\text{int}}\,=\,\frac{\lambda_{1}}{2 \Lambda}\,\left(\bar{e}\PLH F_{ 1}\right)\,\varphi^*\varphi^*\,+\,\frac{\lambda_{2}}{2 \Lambda}\,\left(\bar{e}\PLH F_{ 2}\right)\,\varphi^*\varphi^*\,+\,\frac{\kappa}{\Lambda^2}\,\left(\bar{e}\PLH F_{ 1}\right)\left(\bar{F}_{2}\PRH e\right)\,+\,\text{H.c.}
\end{equation}
where $e$ denotes a SM charged lepton. Such interactions can be motivated by assuming that some heavy scalar mediator, connecting pairs of DM particles with a SM lepton along with an exotic one, has been integrated out. DM stability can be ensured by imposing a $\mathbb{Z}_3$ symmetry under which the $F_i$'s and $\varphi$ are charged according to the assignments summarized in Table \ref{tab:d5_quantum_numbers}. 

The heavy fermions $F_i$ carry the same lepton number as the SM leptons, so their interactions conserve the total lepton number. They are maintained in thermal equilibrium with the thermal plasma due to their gauge interactions [which we have omitted in \Eqs{eq:Lagdim5} for the sake of brevity].

\begin{figure}[t]
    \centering
\begin{subfigure}{.46\linewidth}
  \includegraphics[width=\linewidth]{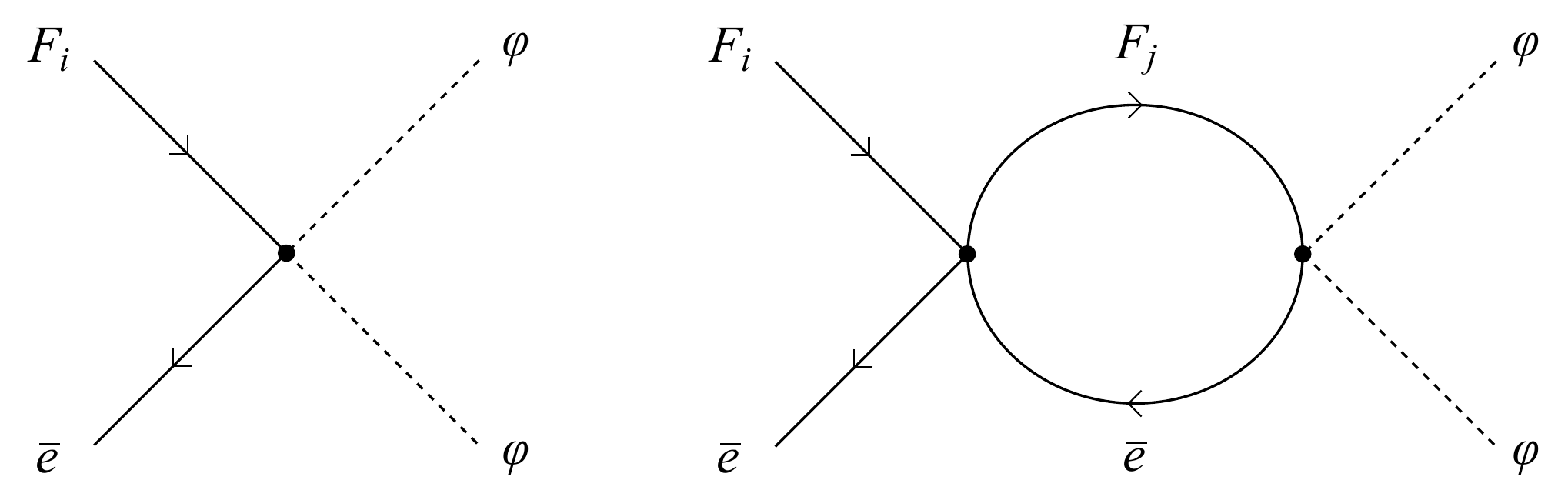}
  \caption{}
  \label{fig:d5_Feynman_diagrams_CP_asymmetry_scattering_1}
\end{subfigure}\hspace*{\fill}
\begin{subfigure}{.46\linewidth}
  \includegraphics[width=\linewidth]{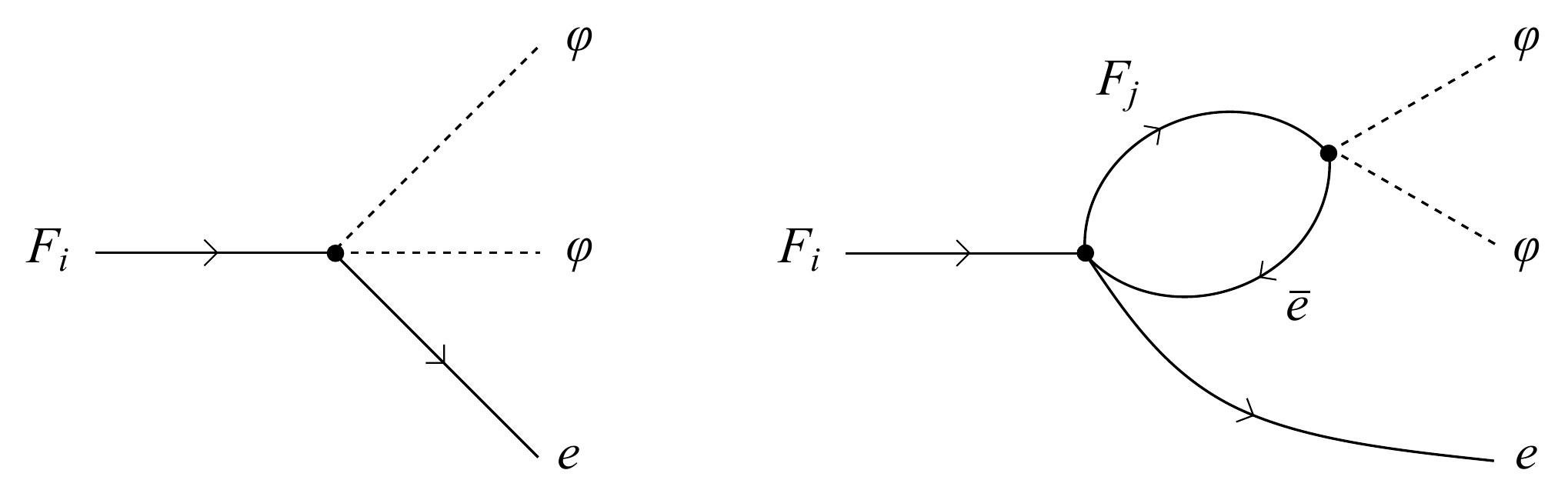}
  \caption{}
  \label{fig:d5_Feynman_diagrams_CP_asymmetry_decay_1}
\end{subfigure}

\medskip

\begin{subfigure}{.46\linewidth}
  \includegraphics[width=\linewidth]{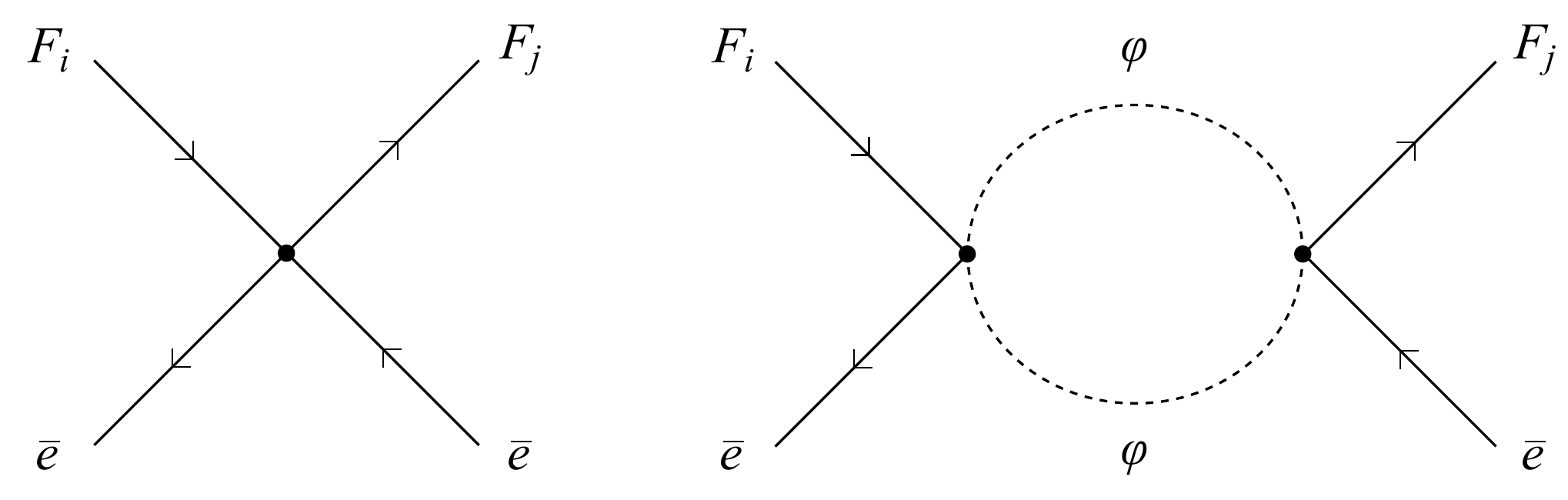}
  \caption{}
  \label{fig:d5_Feynman_diagrams_CP_asymmetry_scattering_2}
\end{subfigure}\hspace*{\fill}
\begin{subfigure}{.46\linewidth}
  \includegraphics[width=\linewidth]{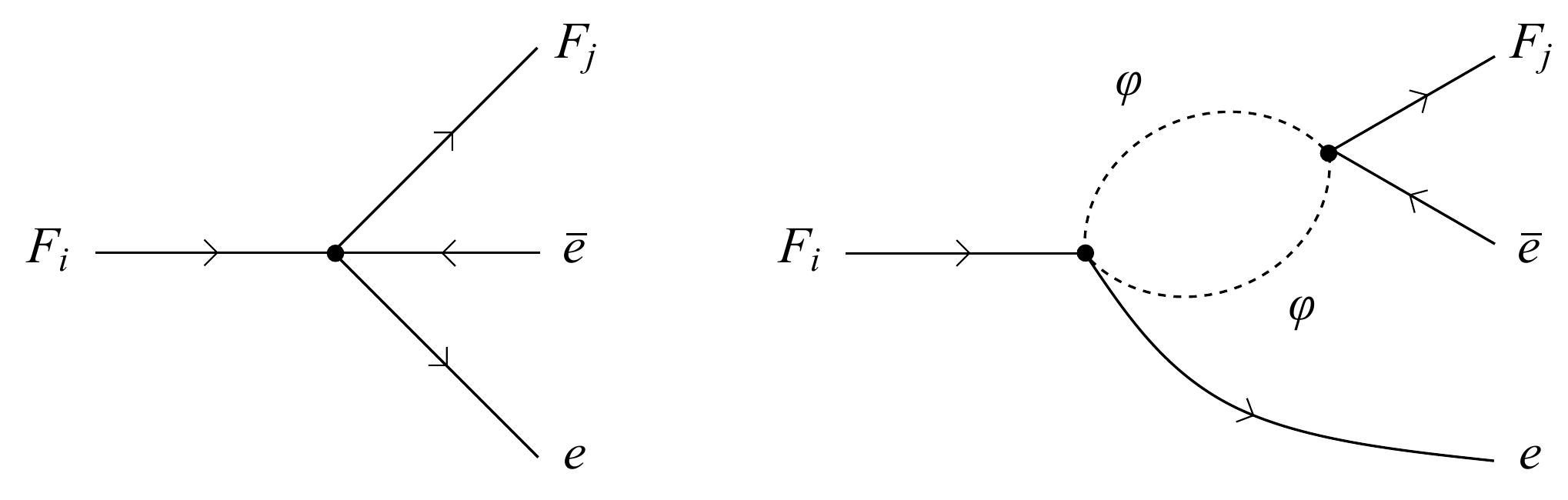}
  \caption{}
  \label{fig:d5_Feynman_diagrams_CP_asymmetry_decay_2}
\end{subfigure}
\caption{Feynman diagrams that contribute to the generation of the baryon asymmetry and DM: (a) pair production of $\varphi$ through scattering, (b) pair production of $\varphi$ through decay, (c) scattering of $F_i$ to $F_j$ and (d) decay of $F_i$ to $F_j$ along with a pair of SM leptons.}
\label{fig:d5_Feynman_diagrams_CP_asymmetry}
\end{figure}

All relevant Feynman diagrams which can contribute to the generation of the baryon asymmetry and DM are shown in Figure \ref{fig:d5_Feynman_diagrams_CP_asymmetry}. The interactions which contain a DM state as both initial and final states are subleading, from the freeze-in assumption, and can be safely ignored.

\begin{table}[h]
    \centering
    \begin{tabular}{l l l}\hline\hline
         Particle & Gauge & $\mathbb{Z}_3$\\
         \hline
         $\varphi$     & $(1,1)_{0}$ & $\omega$ \\
         $\varphi^*$     & $(1,1)_{0}$ & $\omega^{-1}$ \\
         $F_i$         & $(1,1)_{-1}$ & $\omega^{-1}$ \\
         $\bar{F}_i$   & $(1,1)_{1}$ & $\omega$ \\
        \hline\hline
    \end{tabular}
    \caption{Quantum charges of the various states, where $\omega=e^{i2\pi/3}$.}
    \label{tab:d5_quantum_numbers}
\end{table}

Let us first focus on $2 \leftrightarrow 2$ processes. The equilibrium interaction rate density of a generic scattering process $ab\rightarrow cd$ is given by
\begin{align}\label{eq:scattering rate density}
\gamma_{ab\rightarrow cd}\,&\equiv\,\int\text{d}\Pi_{a}  \, \text{d}\Pi_{b} \, \text{d}\Pi_{c} \, \text{d}\Pi_{d}\,\left(2\pi\right)^4\,\delta^{(4)}\left(p_{a}+p_{b}-p_{c}-p_{d}\right)\,f^{\text{eq}}_{a}\,f^{\text{eq}}_{b}\,\left|\mathcal{M}\right|^2_{ab\rightarrow cd}\nonumber
\\
&=\,\frac{T}{512\pi^6}\,\int\text{d}\tilde{s}\,\frac{1}{\sqrt{\tilde{s}}}\,\left|\textbf{p}_{\text{in}}\right|\,\left|\textbf{p}_{\text{fin}}\right|\,K_1\left(\frac{\sqrt{\tilde{s}}}{T}\right)\int\,\text{d}\Omega\,\left|\mathcal{M}\right|^2_{ab\rightarrow cd}
\end{align}
where $\text{d}\Pi_k=\text{d}^3\textbf{p}_k/(\left(2\pi\right)^32E_k)$ is the elementary Lorentz-invariant phase space volume of species $k$, $f_k^{\text{eq}}$ is the equilibrium distribution function of $k$ which we assume to approximately follow the Maxwell-Boltzmann distribution with vanishing chemical potential, $\left|\mathcal{M}\right|^2$ denotes the squared matrix element summed over the internal degrees of freedom of the particles involved, but not averaged over the internal degrees of freedom of the initial states, $K_1$ is the modified Bessel function of the second kind of order one, $\tilde{s}_{\text{min}}=\text{max}\big\{\left(m_a+m_b\right)^2,\left(m_c+m_d\right)^2\big\}$, $\tilde{s}_{\text{max}}\sim\Lambda^2$ and $\textbf{p}_{\text{in}}$, $\textbf{p}_{\text{fin}}$ are the initial and final momenta in the center of momentum (CoM) frame, respectively. An analytical approximation for the scattering rate densities at tree-level, hereafter denoted by $\gamma^{ab}_{cd}$, can be obtained in the limit $\Lambda\gg m_i$,
\begin{subequations}\label{eq:tree-level analytical rate densities model 1}
\begin{align}
\gamma^{F_i\bar{e}}_{\varphi\varphi}\,&\approx\,\frac{\left|\lambda_i\right|^2T}{512\,\pi^5\,\Lambda^2}\int_{0}^{\infty}\text{d}\tilde{s}\,\,\tilde{s}^{3/2}\,K_1\left(\frac{\sqrt{\tilde{s}}}{T}\right) 
\, =\, \frac{\left|\lambda_i\right|^2}{16\,\pi^5}\frac{T^6}{\Lambda^2}
\\\nonumber\\
\gamma^{F_i\bar{e}}_{F_j\bar{e}}\,&\approx\,\frac{\left|\kappa\right|^2T}{512\,\pi^5\,\Lambda^4}\int_{0}^{\infty}\text{d}\tilde{s}\,\,\tilde{s}^{5/2}\,K_1\left(\frac{\sqrt{\tilde{s}}}{T}\right)
\, =\,\frac{3\left|\kappa\right|^2}{2\,\pi^5}\frac{T^8}{\Lambda^4} \; . \label{eq:tree-level analytical rate density model 1}
\end{align}
\end{subequations}

In what follows we will consider $F_i$ masses close to the TeV range and values of $\Lambda\gtrsim 10^{11}\GeV$ (see Figure \ref{fig:constraints}). The $F_i$ decay width can, in turn, be approximated as
\begin{equation}\label{eq:3bodydim5}
\Gamma_{F_i\rightarrow e\varphi\varphi}\,\approx\,\dfrac{\left|\lambda_{i}\right|^2}{3072 \ \pi^3}\frac{M_i^3}{\Lambda^2},\qquad \Gamma_{F_2\rightarrow F_1e\bar{e}}\,\approx\,\dfrac{\left|\kappa\right|^2}{6144 \ \pi^3}\frac{M_2^5}{\Lambda^4}\,,
\end{equation}
where the latter expression holds assuming $M_{1} \ll M_{2}$. Then, for the parameter ranges that will be of interest to us ($\lbrace \Lambda, T_{\rm RH} \rbrace \gg M_i$), the $1\leftrightarrow 3$ decay processes $F_i\leftrightarrow e\varphi\varphi$ and $F_2\leftrightarrow F_1e\bar{e}$ are extremely suppressed and can be safely ignored. 
Instead, the dominant processes for the generation of the DM abundance and the asymmetries are the $2\leftrightarrow 2$ scatterings $F_i\bar{e}\leftrightarrow\varphi\varphi$ and $F_i\bar{e}\leftrightarrow F_j\bar{e}$ which are, moreover, mostly efficient at high temperatures [\textit{cf} Eqs.~\eqref{eq:tree-level analytical rate densities model 1}]. 

The freeze-in DM abundance in our model can be read-off directly from Eq.~\eqref{eq:YDMT}, substituting $n=1$ and $A=|\lambda_1|^2+|\lambda_2|^2$. That is,

\begin{equation}
Y_{\text{DM}}\left(T\right)\,=\,\frac{45\big(\left|\lambda_1\right|^2+\left|\lambda_2\right|^2\big)M_{Pl}}{16\times 1.66\pi^7g_{\star s}\sqrt{g_{\star\rho}} }\frac{T_{\text{RH}}-T}{\Lambda^2}
\end{equation}
where $g_{\star s}=g_{\star\rho}=113.75$, due to the addition of the two Dirac fermions $F_i$ besides the usual SM degrees of freedom. The DM relic density $\Omega_{\text{DM}}h^2$ can, in turn, be cast into the simple form
\begin{equation}
    \Omega_{\rm DM} h^2 \approx 0.12\times\Big(|\lambda_1|^2+|\lambda_2|^2\Big)
    \Big(\dfrac{1.61 \times 10^{16}~\GeV}{\Lambda}\Big)^2
    \Big(\dfrac{T_{\rm RH}}{2\times 10^{15}~\GeV}\Big) 
    \Big(\dfrac{m_{\varphi}}{10 ~\keV}\Big) \;.
    \label{eq:relic_d5}
\end{equation}

\subsubsection{$CP$ asymmetries}

The $CP$ asymmetry is generated, at leading order, due to the interference of the tree-level and 1-loop diagrams presented in Figure \ref{fig:d5_Feynman_diagrams_CP_asymmetry}. As we mentioned, we focus on  the dominant $2\leftrightarrow 2$ scattering processes depicted in Figures \ref{fig:d5_Feynman_diagrams_CP_asymmetry_scattering_1} and \ref{fig:d5_Feynman_diagrams_CP_asymmetry_scattering_2}.

In order to derive an expression for the relevant $CP$ asymmetry let us first consider a generic $2\leftrightarrow 2$ process, denoted by $ab\leftrightarrow cd$. The $CP$ violation can be parametrized as
\begin{equation}
\epsilon\,=\,\frac{\gamma_{ab\rightarrow cd}\,-\,\gamma_{\bar{a}\bar{b}\rightarrow\bar{c}\bar{d}}}{\gamma_{ab\rightarrow cd}\,+\,\gamma_{\bar{a}\bar{b}\rightarrow\bar{c}\bar{d}}}\,\approx\,\frac{\int\text{d}\tilde{s}\,\frac{1}{\sqrt{\tilde{s}}}\,\left|\textbf{p}_{\text{in}}\right|\,\left|\textbf{p}_{\text{fin}}\right|\,K_1\left(\frac{\sqrt{\tilde{s}}}{T}\right)\,\left[\left|\mathcal{M}\right|^2_{ab\rightarrow cd}\,-\,\left|\mathcal{M}\right|^2_{\bar{a}\bar{b}\rightarrow\bar{c}\bar{d}}\right]}{2\int\text{d}\tilde{s}\,\frac{1}{\sqrt{\tilde{s}}}\,\left|\textbf{p}_{\text{in}}\right|\,\left|\textbf{p}_{\text{fin}}\right|\,K_1\left(\frac{\sqrt{\tilde{s}}}{T}\right)\,\left|\mathcal{M}\right|^2_{ab\rightarrow cd}\big|_0}
\end{equation}
where $\bar{i}$ denotes the $CP$-conjugate state of $i$ and we have used the expression for the scattering rate density given in Eq.~\eqref{eq:scattering rate density}. It is convenient to separate the matrix element $\left.\mathcal{M}\right|_k$, with $k$ being the loop order $\left(k=0,1,\ldots\right)$, into a coupling constant part $c_k$ (denoting a collection of coupling constants) and an amplitude part $\mathcal{A}_k$, \textit{i.e.} $\left.\mathcal{M}\right|_k\equiv\sum_{i=0}^{k}\,c_i\mathcal{A}_i$. To leading order the $CP$ asymmetry can be rewritten as \cite{Davidson:2008bu}
\begin{equation}\label{eq:epsgeneral}
\epsilon\,=\,-2\,\frac{\text{Im}\big\{c_0^*c_1\big\}}{\left|c_0\right|^2}\frac{\int\text{d}\tilde{s}\;\frac{1}{\sqrt{\tilde{s}}}\,\left|\textbf{p}_{\text{in}}\right|\,\left|\textbf{p}_{\text{fin}}\right|\,K_1\left(\frac{\sqrt{\tilde{s}}}{T}\right)\,\text{Im}\big\{\mathcal{A}_0^*\mathcal{A}_1\big\}}{\int\text{d}\tilde{s}\,\frac{1}{\sqrt{\tilde{s}}}\,\left|\textbf{p}_{\text{in}}\right|\,\left|\textbf{p}_{\text{fin}}\right|\,K_1\left(\frac{\sqrt{\tilde{s}}}{T}\right)\,\left|\mathcal{A}_0\right|^2}
\end{equation}
where $\text{Im}\big\{\mathcal{A}_0^*\mathcal{A}_1\big\}=\frac{1}{2i}\mathcal{A}_0^*\sum_{\text{cuts}}\mathcal{A}_1$. In the model under consideration there are three $CP$ asymmetries generated by scattering processes, which can be defined as
\begin{subequations}
\begin{alignat}{2}
\epsilon_i\,&\equiv\frac{\gamma_{F_i\bar{e}\rightarrow \varphi\varphi}\,-\,\gamma_{\bar{F}_i e\rightarrow\varphi^*\varphi^*}}{\gamma_{F_i\bar{e}\rightarrow \varphi\varphi}\,+\,\gamma_{\bar{F}_i e\rightarrow\varphi^*\varphi^*}}\label{eq:CP asymmetries 1,2 model 1}
\\\nonumber\\
\epsilon_3\,&\equiv\frac{\gamma_{F_1\bar{e}\rightarrow F_2\bar{e}}\,-\,\gamma_{\bar{F}_1 e\rightarrow\bar{F}_2 e}}{\gamma_{F_1\bar{e}\rightarrow F_2\bar{e}}\,+\,\gamma_{\bar{F}_1 e\rightarrow\bar{F}_2 e}}\label{eq:CP asymmetry 3 model 1}
\end{alignat}
\end{subequations}
where $i=1,2$. Let us note that whereas the processes corresponding to the $\epsilon_i$ quantities generate \textit{correlated} $CP$ asymmetries in the two sectors (\textit{i.e.} in the SM and the exotic fermions), the reactions of the type $F_1\bar{e}\leftrightarrow F_2\bar{e}$ do not directly affect the asymmetry generated in the SM leptons. Note that when $\epsilon_3$ is embedded in the Boltzmann equations the corresponding source term vanishes, as the external states follow approximately their equilibrium abundances. However, it has to be included in our study in order to consistently use the $CPT$ and unitarity relations. Besides, although in the denominator of \eqref{eq:CP asymmetry 3 model 1} the tree-level and one-loop contributions (\textit{cf} Fig.~\ref{fig:d5_Feynman_diagrams_CP_asymmetry_scattering_2}) both scale as $1/\Lambda^2$, the latter is suppressed by an additional loop factor and can, thus safely be ignored (as long as $\kappa \sim \lambda_{1,2}$, an assumption which we will be adopting throughout our numerical analysis).

The imaginary part of the amplitudes entering the $\epsilon_i$ asymmetries can be calculated from the optical theorem to be (see Appendix \ref{sec:Appendix})
\begin{equation}\label{eq:imidim5}
\text{Im}\big\{\mathcal{A}_0^*\mathcal{A}_1\big\}_{F_i\bar{e}\rightarrow\varphi\varphi}\,=\,\frac{\big(\tilde{s}-M_i^2-m_e^2\big)\big(\tilde{s}-M_j^2-m_e^2\big)\,\sqrt{\lambda\big(\tilde{s},M_j^2,m_{e}^2\big)}}{16\pi\tilde{s}}
\end{equation}
where $\lambda$ is the K{\"a}ll{\'e}n (or triangle) function. This expression has been crossed checked using {\tt Package-X} \cite{Patel:2015tea}, as well as by implementing the Lagrangian of Eq.~\eqref{eq:Lagdim5} in {\tt FeynRules} \cite{Alloul:2013bka} and then performing the computation through {\tt Feynarts/FormCalc} \cite{Hahn:2000kx,Hahn:1998yk}. 

An analytical estimation for the $\epsilon_1$ and $\epsilon_2$ $CP$ asymmetries can be obtained in the limit $\Lambda\gg m_i$, yielding
\begin{subequations}\label{eq:epsidim5}
\begin{align}
\epsilon_1\,&\approx\,-\frac{3}{\pi}\frac{\left|\kappa\right|\left|\lambda_2\right|}{\left|\lambda_1\right|}\sin\left(-\Delta \phi\right)\frac{T^2}{\Lambda^2}
\\\nonumber\\
\epsilon_2\,&\approx\,-\frac{3}{\pi}\frac{\left|\kappa\right|\left|\lambda_1\right|}{\left|\lambda_2\right|}\sin\left(\Delta \phi \right)\frac{T^2}{\Lambda^2}
\end{align}
\end{subequations}
where we have expressed the coupling constants in terms of their magnitude and phase, \textit{i.e.} $\lambda_i=|\lambda_i|e^{i\phi_i}$ and $\kappa=|\kappa|e^{i\phi_3}$, and we have 
defined $\Delta \phi = \phi_1-\phi_3-\phi_2$. In the subsequent analysis, we will set $\Delta\phi = \pi/2$ in order to maximize the contribution of the phases to the $CP$ violation.

In the same way we can calculate the imaginary part of the Feynman amplitudes for the case of the $\epsilon_3$ asymmetry to be
\begin{equation}\label{eq:im3dim5}
\text{Im}\big\{\mathcal{A}_0^*\mathcal{A}_1\big\}_{F_1\bar{e}\rightarrow F_2\bar{e}}\,=\,\frac{\big(\tilde{s}-M_1^2-m_e^2\big)\big(\tilde{s}-M_2^2-m_e^2\big)\,\sqrt{\lambda\big(\tilde{s},m_{\varphi}^2,m_{\varphi}^2\big)}}{16\pi\tilde{s}} \; .
\end{equation}
In the limit $\Lambda\gg m_i$  we can, again, derive an analytical approximation for $\epsilon_3$ as
\begin{equation}\label{eq:eps3approx}
\epsilon_3\,\approx\,-\frac{1}{16\pi}\frac{\left|\lambda_1\right|\left|\lambda_2\right|}{\left|\kappa\right|}\sin\left(\Delta \phi \right) \; .
\end{equation}
Contrary to $\epsilon_i$ case the $\epsilon_3$ asymmetry is constant during the cosmic expansion, which can be understood by dimensional arguments noting that the combination of the couplings that enters $\epsilon_3$ eliminates the EFT energy scale $\Lambda$.

\begin{figure}[t]
\centering
\includegraphics[width=0.7\linewidth]{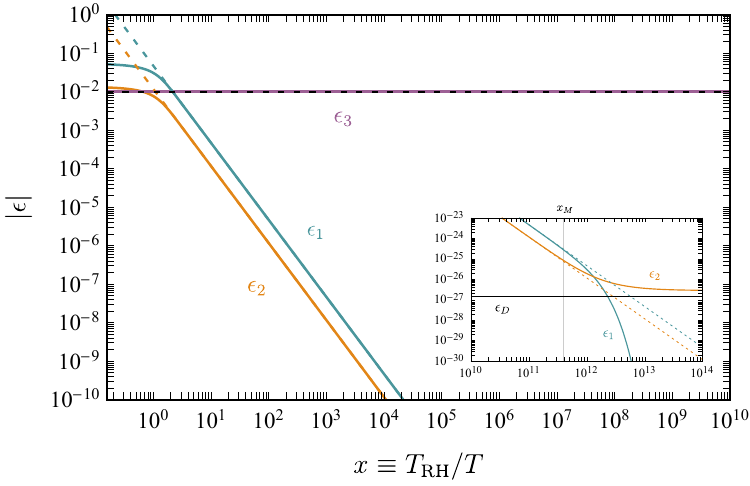}
\caption{Temperature evolution of the $CP$ asymmetries assuming $\Lambda=1.28\times 10^{16}\GeV$, $T_{\text{RH}}=2\times10^{15}\GeV$, $M_1=1\TeV$, $M_2=5\TeV$, $|\lambda_1|=0.5$, $|\lambda_2|=|\kappa|=1$ and $\Delta\phi=\pi/2$. The colored solid lines correspond to the full numerical results, while the dashed lines to the analytical approximations as described in the text. The inset focuses on the evolution of the $\epsilon$'s at low temperatures and the horizontal black solid line corresponds to the $CP$ asymmetry generated by the decay $F_2\rightarrow e\varphi\varphi$.}
\label{fig:CP Asymmetries model 1}
\end{figure}

In Figure \ref{fig:CP Asymmetries model 1} we illustrate the $CP$ asymmetries predicted by our model as a function of the dimensionless parameter $x\equiv T_{\text{RH}}/T$, for a benchmark set of parameter values: $\Lambda=1.28\times10^{16}\GeV$, $T_{\text{RH}}=2\times10^{15}\GeV$, $M_1=1\TeV$, $M_2=5\TeV$, $|\lambda_1|=0.5$, $|\lambda_2|=|\kappa|=1$ and $\Delta\phi=\pi/2$, which, as we will see shortly, lead to the observed baryon asymmetry of the Universe. The solid lines correspond to the full numerical results obtained through Eqs.~\eqref{eq:epsgeneral}, \eqref{eq:imidim5} and \eqref{eq:im3dim5}, whereas the dashed lines to the analytical approximations of Eqs.~\eqref{eq:epsidim5} and \eqref{eq:eps3approx}. In the temperature range which is relevant for our study ($T<T_{\rm RH}$), we observe that the numerical results are in very good agreement with the analytical ones. Besides, as the temperature drops below the mass $M_2$ of the heavier fermion, \textit{i.e.} below $x_M\equiv T_{\text{RH}}/M_2$, the $\epsilon_1$ asymmetry becomes kinematically suppressed, as there is not enough energy for the $F_2$ fermion to propagate on-shell in the loop. On the other hand, $\epsilon_2$ takes values close to the $CP$ asymmetry generated by the heavier fermion decay $F_2 \rightarrow e\phi\phi$, which can be approximated as $\epsilon_D\approx\frac{\left|\lambda_2\right|^2}{32\pi}\frac{M_2^2}{\Lambda^2}$.

Before setting off to compute the baryon asymmetry of the Universe predicted by our model let us, finally, note that unitarity of the $S$-matrix together with invariance under $CPT$ impose specific relations between the various asymmetries $\epsilon_{1,2,3}$ and the corresponding equilibrium reaction rate densities. For a generic process $a_i\leftrightarrow b_j$, these two conditions imply that \cite{Kolb:1979qa}
\begin{equation}
\sum_{b_j,\bar{b}_j}\gamma_{a_i\rightarrow b_j}\,=\, \sum_{b_j,\bar{b}_j}\gamma_{b_j\rightarrow a_i}\,=\, \sum_{b_j,\bar{b}_j}\gamma_{\bar{a}_i\rightarrow \bar{b}_j}\,=\, \sum_{b_j,\bar{b}_j}\gamma_{\bar{b}_j\rightarrow \bar{a}_i}
\end{equation}
where  $a_i$ and $b_j$ denote a set of states with $i=1,\dots,N$ and $j=1,\dots,M$ and the bar denotes the $CP$-conjugate states. Hence, $CPT$ and unitarity ensure that if thermal equilibrium is attained, then, no asymmetry can be produced. In our model, these requirements translate into the following conditions, which we will apply throughout our subsequent analysis
\begin{subequations}\label{eq:CPT+unitarity conditions model 1}
\begin{align}
\epsilon_1\gamma^{F_1\bar{e}}_{\varphi\varphi}\,+\,\epsilon_3\gamma^{F_1\bar{e}}_{F_2\bar{e}}\,&=\,0
\\\nonumber\\
\epsilon_2\gamma^{F_2\bar{e}}_{\varphi\varphi}\,-\,\epsilon_3\gamma^{F_1\bar{e}}_{F_2\bar{e}}\,&=\,0
\\\nonumber\\
\epsilon_1\gamma^{F_1\bar{e}}_{\varphi\varphi}\,+\,\epsilon_2\gamma^{F_2\bar{e}}_{\varphi\varphi}\,&=\,0 \; .
\end{align}
\end{subequations}
Moreover, the reaction $F_1\bar{F}_2\leftrightarrow e\bar{e}$, as well as all gauge interactions, are $CP$-conserving which is the reason why, with some amount of hindsight, we have neglected them in our discussion.

\subsubsection{Baryon asymmetry}\label{sec:baryondim5}

Having discussed how $C/CP$ violation and out-of-equilibrium dynamics arise in the model described by the Lagrangian of Eq.~\eqref{eq:Lagdim5}, let us now turn to baryon (or lepton) number violation. 

First, note that in our model the heavy fermions $F_i$ are of Dirac nature and carry the same lepton number as the SM leptons. Then, once sector-wise asymmetries are generated in the SM and exotic fermion sectors, the total lepton asymmetry is given by $Y_L=Y_{L_{\text{SM}}}+Y_{L_F}$, where $Y_{L_F} \equiv \sum_iY_{L_{F_i}}$. On the other hand, all processes (including non-perturbative sphaleron transitions) conserve the combination $Y_{B-L}\equiv Y_B-Y_{L_{\text{SM}}}-Y_{L_F}$, \textit{i.e.} $\text{d}Y_{B-L}/\text{d}T=0$. If we also assume that the Universe is initially symmetric, $Y_{B-L_{\text{SM}}}=Y_{L_F}=0$, then at any $T$ it holds that $Y_{B-L_{\text{SM}}}=Y_{L_F}$, \textit{i.e.} equal and opposite lepton asymmetries are generated in the two sectors and the net lepton/baryon asymmetry can only be zero.

The situation changes once sphaleron transitions are taken into account. Whereas sphalerons are insensitive to the heavy lepton asymmetry $Y_{L_F}$, as the $F_i$ are $SU(3)_c\times SU(2)_{\LH}$ singlets, they only affect the non-zero lepton asymmetry $Y_{L_{\text{SM}}}$ stored in the SM sector. In particular, they convert it into a baryon asymmetry by imposing certain relations among the chemical potentials of the various species. Once sphalerons depart from equilibrium, which occurs at $T_{sph}=131.7\pm 2.3 \GeV$ \cite{DOnofrio:2014rug}, the baryon and lepton numbers are separately conserved. In principle, when the relic heavy leptons eventually decay away the net baryon asymmetry, being proportional to $Y_{B-L_{\text{SM}}}$, would vanish. However, if the sphalerons are inactive during the decay epoch of $F_i$ (which, as we will see in Section \ref{sec:resultsdim5}, is the case throughout our viable parameter space), then the baryon asymmetry remains frozen at a value $Y_B\propto Y_{B-L_{\text{SM}}}|_{T_{sph}}$, which, in general, is non-zero.

Taking into consideration only the $2\leftrightarrow 2$ scatterings and ignoring the subleading $1\leftrightarrow 3$ decay processes, the Boltzmann equations of the asymmetries read
\begin{subequations}
\begin{align}
-sHT\frac{\text{d}Y_{\Delta F_i}}{\text{d}T}\,&=\,-\big[F_i\bar{e}\leftrightarrow\varphi\varphi\big]\,-\,\big[F_i\bar{e}\leftrightarrow F_j\bar{e}\big]\,+(-1)^i\,\big[F_i\bar{F}_j\leftrightarrow e\bar{e}\big]
\\\nonumber\\
-sHT\frac{\text{d}Y_{B-L_{\text{SM}}}}{\text{d}T}\,&=\,-\sum_i\big[F_i\bar{e}\leftrightarrow\varphi\varphi\big]
\end{align}
\end{subequations}
where $Y_{\Delta F_i}\equiv Y_{F_i}-Y_{\bar{F}_i}$ and we have used the notations,
\begin{subequations}\label{eq:notation conventions}
\begin{alignat}{2}
\left[a\,b\leftrightarrow c\,d\right]\,&\equiv\,\left(a\,b\leftrightarrow c\,d\right)\,-\,\left(\bar{a}\,\bar{b}\leftrightarrow \bar{c}\,\bar{d}\right)
\\
\left(a\,b\leftrightarrow c\,d\right)\,&\equiv\,\frac{1}{S_{\text{in}}S_{\text{out}}}\int\text{d}\Pi_a \, \text{d}\Pi_b \, \text{d}\Pi_c \, \text{d}\Pi_d \, \left(2\pi\right)^4\delta^{(4)}\Big[\left|\mathcal{M}\right|^2_{ab\rightarrow cd}f_af_b\left(1\pm f_c\right)\left(1\pm f_d\right)\nonumber
\\
&\quad\qquad\qquad\qquad\qquad\qquad\qquad\qquad\qquad\qquad-\,\left|\mathcal{M}\right|^2_{cd\rightarrow ab}f_cf_d\left(1\pm f_a\right)\left(1\pm f_b\right)\Big]
\end{alignat}
\end{subequations}
with $\delta^{\left(4\right)}$ being an abbreviation for $\delta^{\left(4\right)}\left(p_a+p_b-p_c-p_d\right)$ and $S_{\text{in}}$, $S_{\text{out}}$ are the symmetry factors for the incoming and outgoing states respectively, which are equal to $N!$ in case of identical particles and $1$ otherwise. We further approximate the distribution functions of the visible sector species by Maxwell-Boltzmann statistics, \textit{i.e.} we neglect Bose-enhancement and Pauli-blocking factors. Hence, we can write
\begin{equation}
\left(a\,b\leftrightarrow c\,d\right)\,=\,\frac{1}{S_{\text{in}}S_{\text{out}}}\bigg(\gamma_{ab\rightarrow cd}\,\frac{Y_a}{Y_a^{\text{eq}}}\,\frac{Y_b}{Y_b^{\text{eq}}}\,-\,\gamma_{cd\rightarrow ab}\,\frac{Y_c}{Y_c^{\text{eq}}}\,\frac{Y_d}{Y_d^{\text{eq}}}\bigg) \; .
\end{equation}
Let us also point out that throughout the subsequent analysis we will ignore thermal corrections to the masses of the particles involved in the various processes. Since these corrections should affect in a similar manner the masses of the $F_i$'s and the SM leptons, we expect them not to qualitatively alter our results.

The various terms in the Boltzmann equation can be expressed in terms of the $CP$ asymmetries, the tree-level scattering rate densities and the asymmetric abundances. Due to the small value of the observed baryon asymmetry, it is typical to linearize in the SM chemical potentials \cite{Kolb:1979qa}
\begin{equation}\label{eq:linearize in chemical potentials}
\frac{Y_{e(\bar{e})}}{Y_e^{\text{eq}}}\,\equiv\,e^{\pm\mu_e/T}\,\approx\,1\,\pm\,\frac{\mu_e}{T}\,=\,1\,\pm\,\frac{Y_{\Delta e}}{2Y_{\gamma}}
\end{equation}
Using \eqref{eq:linearize in chemical potentials} as well as the $CPT$ and unitarity conditions of Eqs.~\eqref{eq:CPT+unitarity conditions model 1}, the Boltzmann equations can be rewritten as
\begin{subequations}
\begin{align}
& -sHT \frac{\text{d}Y_{\Delta F_i}}{\text{d}T}\,=\,\epsilon_i\gamma^{F_i\bar{e}}_{\varphi\varphi}\left[\frac{Y_{F_j+\bar{F}_j}}{Y_{F_j+\bar{F}_j}^{\text{eq}}}\,-\,\left(\frac{Y_{\varphi}}{Y_{\varphi}^{\text{eq}}}\right)^2\right]-\frac{1}{2}\gamma^{F_i\bar{e}}_{\varphi\varphi}\left(y_{F_i}-y_e\right) +
\nonumber\\
&+(-1)^i\,\gamma^{F_1\bar{e}}_{F_2\bar{e}}\left(y_{F_1}-\frac{Y_{F_1+\bar{F}_1}}{Y_{F_1+\bar{F}_1}^{\text{eq}}}\,y_e-y_{F_2}+\frac{Y_{F_2+\bar{F}_2}}{Y_{F_2+\bar{F}_2}^{\text{eq}}}\,y_e\right)\,+(-1)^i\,\gamma^{F_1\bar{F}_2}_{e\bar{e}}\left(\frac{Y_{F_1}}{Y_{F_1}^{\text{eq}}}\frac{Y_{\bar{F}_2}}{Y_{F_2}^{\text{eq}}}\,-\,\frac{Y_{\bar{F}_1}}{Y_{F_1}^{\text{eq}}}\frac{Y_{F_2}}{Y_{F_2}^{\text{eq}}}\right)\label{eq:BEF_initial}
\\\nonumber\\
& -sHT\frac{\text{d}Y_{B-L_{\text{SM}}}}{\text{d}T}\, = \,\epsilon_1\gamma^{F_1\bar{e}}_{\varphi\varphi}\left(\frac{Y_{F_2+\bar{F}_2}}{Y_{F_2+\bar{F}_2}^{\text{eq}}}-\frac{Y_{F_1+\bar{F}_1}}{Y_{F_1+\bar{F}_1}^{\text{eq}}}\right)\,-\,\frac{1}{2}\gamma^{F_1\bar{e}}_{\varphi\varphi}\left(y_{F_1}-y_e\right)\,-\,\frac{1}{2}\gamma^{F_2\bar{e}}_{\varphi\varphi}\left(y_{F_2}-y_e\right) \label{eq:BEL_initial}
\end{align}
\end{subequations}
where $y_{F_i}\equiv Y_{\Delta F_i}/Y_{F_i}^{\text{eq}}$ and $y_e\equiv Y_{\Delta e}/Y_{\gamma}$. Note that we have used the $CPT$ and unitarity conditions in order to show explicitly that the source term of each asymmetry vanishes when all species follow their equilibrium abundance. In other words, no asymmetries are generated unless there is a departure from equilibrium, in accordance with the third Sakharov condition. 

Under the freeze-in assumption $Y_{\varphi}/Y_{\varphi}^{\text{eq}}\approx 0$ and, if we also consider the heavy leptons $F_i$ to be kept close to their equilibrium abundances due to their gauge interactions, the Boltzmann equations of the asymmetries simplify to  
\begin{figure}[t]
\centering
\includegraphics[width=0.7\linewidth]{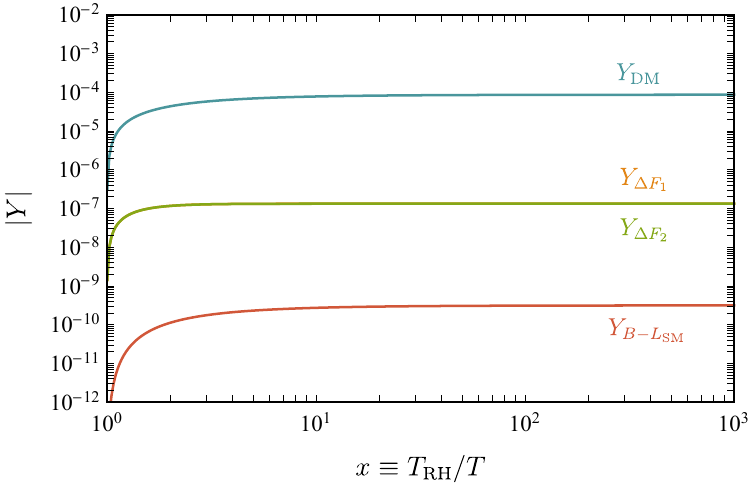}
\caption{The generated asymmetries $Y_{\Delta F_i}, Y_{B-L_{\text{SM}}}$ and the DM abundance $Y_{\text{DM}}$, in terms of the dimensionless parameter $x\equiv T_{\text{RH}}/T$. Observe that the asymmetric yields freeze-in very close to the reheating temperature.}
\label{fig:Asymmetric yields model 1}
\end{figure}
\begin{subequations}\label{eq:d5-BEe}
\begin{align}
-sHT\frac{\text{d}Y_{\Delta F_i}}{\text{d}T}\,&=\,\epsilon_i\gamma^{F_i\bar{e}}_{\varphi\varphi}\,-\,\frac{1}{2}\gamma^{F_i\bar{e}}_{\varphi\varphi}\left(y_{F_i}-y_e\right)\,+(-1)^i\,\gamma^{F_1\bar{e}}_{F_2\bar{e}}\left(y_{F_1}-y_{F_2}\right)
\label{eq:d5-BEFi}\\\nonumber\\
-sHT\frac{\text{d}Y_{B-L_{\text{SM}}}}{\text{d}T}\,&=\,-\frac{1}{2}\gamma^{F_1\bar{e}}_{\varphi\varphi}\left(y_{F_1}-y_e\right)\,-\,\frac{1}{2}\gamma^{F_2\bar{e}}_{\varphi\varphi}\left(y_{F_2}-y_e\right) \;.
\end{align}
\end{subequations}
In passing, let us point out that the first term of Eq.~\eqref{eq:BEL_initial} vanishes, since $F_{1,2}$ are kept in thermal equilibrium due to their gauge interactions. In contrast, the heavy leptons asymmetries have a non-vanishing $CP$-source term and conservation of $Y_{B-L}$ implies that $\sum_i Y_{\Delta F_i}=Y_{B-L_{\text{SM}}}$. 

Due to the fact that the interactions responsible for the creation of all asymmetries are non-renormalizable, the asymmetries freeze-in close to the reheating temperature -- which in our case will turn out to be higher than $10^{13}\GeV$. In this regime all relevant spectator processes which could redistribute the asymmetries initially created in the right-handed SM leptons are inactive \cite{Nardi:2005hs}, which implies that $Y_{\Delta e}=-Y_{B-L_{\text{SM}}}$. We will solve the Boltzmann Eqs.~\eqref{eq:d5-BEe} making use of this observation. At temperatures higher than $10^{12}\GeV$ the electroweak sphalerons are out-of-equilibrium and therefore no baryon asymmetry can be generated. However, as the temperature drops below this value, the sphalerons become active and the baryon asymmetry becomes proportional to $Y_{B-L_{\text{SM}}}$. The proportionality constant depends on all of the spectator processes (\textit{e.g.} Yukawa interactions, sphaleron transitions) which are active and relate the chemical potentials of the participating species \cite{PhysRevD.42.3344,Nardi:2005hs}, including the beyond SM heavy fermions. 
At $T\ll 10^8\GeV$, when all spectator processes are in equilibrium, the baryon asymmetry is related to $Y_{B-L_{\text{SM}}}$ as \cite{Goudelis:2021qla}
\begin{equation}
Y_B\,=\,\frac{22}{79}Y_{B-L_{\text{SM}}} \, .
\end{equation}
The system of the three Boltzmann equations, Eqs.~\eqref{eq:d5-BEe}, has been solved numerically using both the numerical and the analytical expressions for the various rates and $CP$ asymmetries, with the results being in good agreement. As an illustration, we present an explicit example in Figure \ref{fig:Asymmetric yields model 1}, where the asymmetric yields and the DM abundance (the latter being included for completeness) have been calculated using the same set of parameters as in Fig.~\ref{fig:CP Asymmetries model 1}. These parameters reproduce the observed baryon asymmetry of the Universe, while in order to obtain the observed DM relic density the mass of the DM state, as can be seen from Eq.~\eqref{eq:relic_d5}, has to be set to $m_{\varphi}\approx 5.1\keV$. 

One might note that, strictly speaking, once spectator processes become active, the relation $Y_{\Delta e}=-Y_{B-L_{\text{SM}}}$ no longer holds and rather translates to a proportionality relation, with the proportionality constant depending on the temperature under consideration. For example below $T\sim 10^8\GeV$ the two are related through $Y_{\Delta e}=-(28/79)Y_{B-L_{\text{SM}}}$. We have numerically verified that, at least for the values of the reheating temperature that we will consider, replacing $Y_{\Delta e}$ with $-Y_{B-L_{\text{SM}}}$ in our Boltzmann equations constitutes a good approximation within an accuracy better than $1\%$. This is due to the fact that, as can be seen from Fig.~\ref{fig:Asymmetric yields model 1}, the $B-L_{\text{SM}}$ asymmetry is generated extremely close to the reheating temperature (which, in our example, is much higher than $10^{13}$ GeV). We will, hence, adopt this approximation throughout the rest of our numerical treatment.

\subsubsection{Results}\label{sec:resultsdim5}

\begin{figure}[t]
\hspace*{-1.9cm}
\begin{subfigure}{.6\linewidth}
  \includegraphics[width=\linewidth]{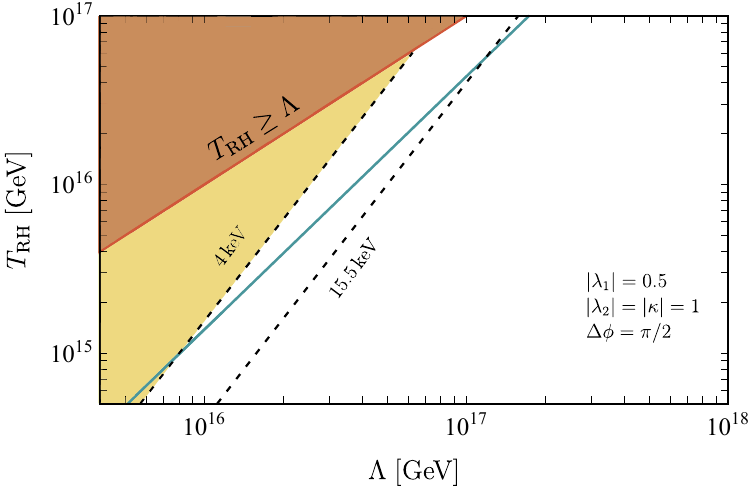}
  \caption{}
  \label{fig:Baryon asymmetry scan model 1}
\end{subfigure}\hspace*{-0.0cm}
\begin{subfigure}{.61\linewidth}
\vspace*{-0.2cm}  
\includegraphics[width=\linewidth]{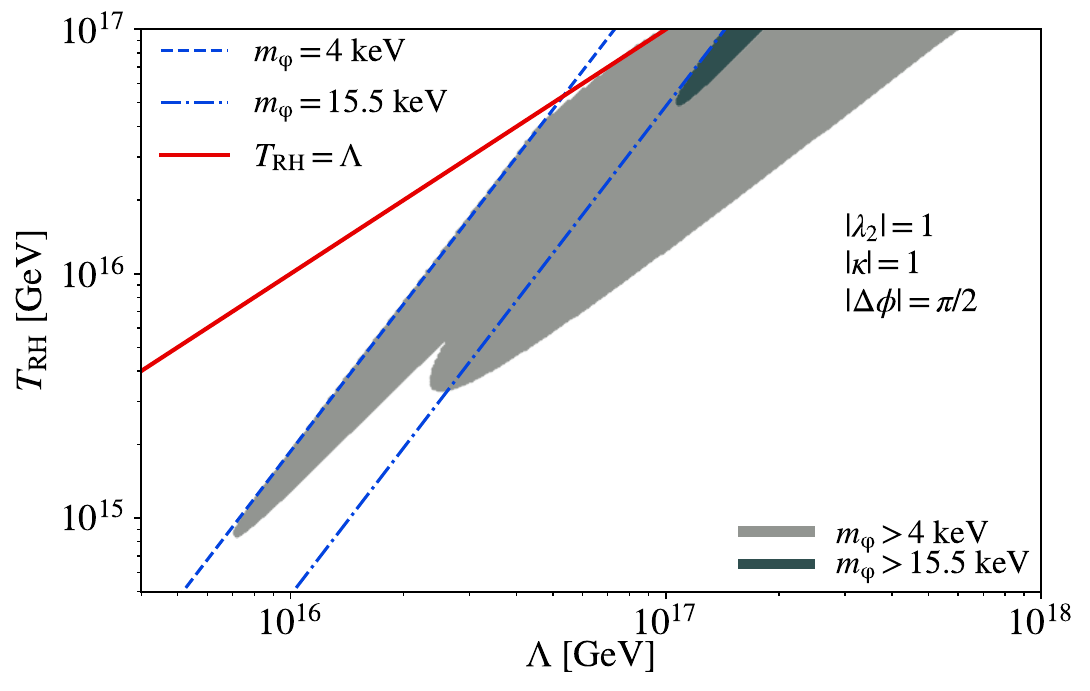}
  \caption{}
  \label{fig:d5_obs}
\end{subfigure}
\caption{(a) Combinations of the reheating temperature $T_{\text{RH}}$ and the energy scale $\Lambda$, which can generate the observed baryon asymmetry. The coupling constants have been set to $|\lambda_1|=0.5$, $|\lambda_2|=|\kappa|=1$ and $\Delta\phi=\pi/2$. The dashed lines depict representative DM masses $m_{\varphi}=\{4,15.5\}\keV$ for which the observed DM relic density can be reproduced. Masses below $4\keV$ are excluded from current Lyman-$\alpha$ forest observations (yellow-shaded area). (b) The values of  $\Lambda$ and $T_{\rm RH}$ that result in the observed DM relic and baryon asymmetry for $|\kappa| = |\lambda_{2}| = 1$ and $\Delta\phi=\pi/2$, while $|\lambda_1|$ is chosen so that $\Omega_{\text{DM}} h^2 =0.12$. The darker-shaded region corresponds to $m_{\varphi} > 15.5~\keV$, and it is extended to the lighter-shaded one for $m_{\varphi} > 4~\keV$. The blue lines show the upper bound on $T_{\rm RH}$ for which the observed DM abundance in the Universe can be reproduced for each value of $\Lambda$, for the two DM mass limits.}
\label{fig:Model_1_res}
\end{figure}

Our results for the predictions of the model described by the Lagrangian of Eq.~\eqref{eq:Lagdim5} are summarized in Figure \ref{fig:Model_1_res}. In the left-hand side panel, we numerically solve the relevant Boltzmann equations, Eqs.~\eqref{eq:d5-BEe}, while adopting the analytical approximations for the $CP$ asymmetries and the scattering rate densities that we presented previously. We show the combinations of the reheating temperature and the EFT energy scale for which the observed baryon asymmetry can be obtained, fixing the relevant couplings at the values: $|\lambda_1|=0.5$, $|\lambda_2|=|\kappa|=1$ and $\Delta\phi=\pi/2$.

In order to extend our scan of the parameter space in an efficient manner, we first note that the third Boltzmann equation admits a formal solution as
\begin{equation}
    Y_{B-L_{\text{SM}}} = C \, e^{\frac{4}{5} \, (|\lambda_1|^2 + |\lambda_2|^2) \, C \, T}
    \int_{T_{\rm RH}}^T d T^{\prime} \, e^{-\frac{4}{5} \, (|\lambda_1|^2 + |\lambda_2|^2) \, C \, T^{\prime}}
    \lrb{|\lambda_1|^2 Y_{\Delta F_1}+ |\lambda_2|^2 Y_{\Delta F_2}}\;,
    \label{eq:d5-YDe}
\end{equation}
where
\begin{equation}
C = \sqrt{\frac{3}{2}} \frac{45}{128 \ \pi^{15/2}}   \frac{M_{Pl}}{ \Lambda^2 } \frac{1}{Y_{F_{1,2}}^{\rm eq} g_{\star s}\sqrt{g_{\star\rho}}}\Big|_{T=T_{\rm RH}} \;.
\end{equation}
Moreover, an estimate of $Y_{\Delta F_i}$ can be obtained by keeping only the first terms in the \rhs of Eqs.~\eqref{eq:d5-BEFi}. The reason for this is that while the sector-wise asymmetries are produced close to the reheating temperature (\textit{cf} the previous discussion and Fig.~\ref{fig:Asymmetric yields model 1}), the second and third terms become active at later times when the yields become sizeable; however, by this time the interaction rates have already become considerably smaller, since they drop quickly with the temperature. Therefore, 
\begin{equation}
    Y_{\Delta F_i} \approx -\int_{T_{\text{RH}}}^T d T^{\prime} \, \epsilon_i\, \dfrac{\gamma^{F_i\bar{e}}_{\varphi\varphi}}{HsT^{\prime}}\;.
    \label{eq:d5-YF_approx}
\end{equation}

By substituting Eq.~\eqref{eq:d5-YF_approx} into Eq.~\eqref{eq:d5-YDe}, expanding for $\Lambda \gg T_{\rm RH}$ and keeping only the leading term, we obtain
\begin{align} \label{eq:d5-YDe_approx}
    \frac{22}{79}Y_{B-L_{\text{SM}}}  \approx  -  \, 8.71\times 10^{-11} \,
    |\kappa|  |\lambda_1| |\lambda_2| \lrb{|\lambda_1|^2-|\lambda_2|^2}
    \sin\lrb{\Delta \phi} \times & \\ \nonumber
    \times \lrb{\dfrac{T_{\rm RH}}{3.3 \times 10^{16}~\GeV}}^{4}
    \lrb{\dfrac{10^{17}~\GeV}{\Lambda}}^{6} & .
\end{align}
As expected from the $\epsilon_i$ $CP$ asymmetries in Eqs.~\eqref{eq:epsidim5}, the asymmetry in the SM lepton vanishes if the couplings $\lambda_{1,2}$ are equal, since the production of lepton asymmetry from both $F_{1,2}$ becomes equal and opposite. 

By requiring the DM relic abundance and the baryon asymmetry to match their corresponding observed values, $\Omega_{\text{DM}} h^2 \approx 0.12$ and $Y_{B} \approx 8.71 \times 10^{-11}$, and using Eqs.~\eqref{eq:relic_d5} and \eqref{eq:d5-YDe_approx}, we find that the cut-off and the reheating temperature have to satisfy
\begin{subequations}\label{eq:d5_obs}
\begin{align}
    & \Lambda \approx 2 \times 10^{16}~\GeV \ 
    \dfrac{\lrb{|\lambda_1|^2+|\lambda_2|^2}^2}{\sqrt{\left||\kappa||\lambda_1||\lambda_2|^2(|\lambda_1|^2-|\lambda_2|^2)\sin\lrb{\Delta \phi}  \right|}} 
    \lrb{\dfrac{m_{\varphi}}{10~\keV}}^2
    \\\nonumber\\
    &T_{\rm RH} \approx 3 \times 10^{15} ~\GeV \ 
    \dfrac{\lrb{|\lambda_1|^2+|\lambda_2|^2}^3}{\left||\kappa||\lambda_1||\lambda_2|^2(|\lambda_1|^2-|\lambda_2|^2)\sin\lrb{\Delta \phi}  \right|} 
    \lrb{\dfrac{m_{\varphi}}{10~\keV}}^3  \; .
\end{align}
\end{subequations}
Scanning over the parameter space, both numerically and analytically using Eqs.~\eqref{eq:d5_obs}, we obtain the allowed values of $\Lambda$ and $T_{\rm RH}$ delineated in Figure~\ref{fig:d5_obs} (shaded regions).

A few comments are in order. First, as we discussed in Section \ref{sec:baryondim5}, in order for the generated baryon asymmetry to survive until the present-day era, the decay of the relic $F_i$'s has to occur after sphaleron decoupling. By employing Eq.~\eqref{eq:3bodydim5} we find this to indeed be the case within the entire viable parameter space depicted in Figure \ref{fig:Model_1_res}. In particular, we find that if the mass of the heavy fermions is smaller than a few TeV, they tend to decay during the big bang nucleosynthesis (BBN) era or even later. At the same time, the (temporary, \textit{i.e.} prior to their eventual decay) relic abundance of the $F_i$'s is comparable to that of thermal WIMPs, since they interact electromagnetically. Based on the analysis performed, \textit{e.g.}, in \cite{Kawasaki:2017bqm}, particles of such abundances decaying into leptons should possess a lifetime shorter than $\sim 100$ seconds. In our case this constraint can, however, be easily satisfied either by simply requiring the masses of the $F_i$'s to be larger than $\sim 5$ TeV or by introducing an additional (faster) decay channel. This additional channel, does not change our freeze-in picture, as its mass -- and, in turn, its contribution to the DM relic abundance -- can be assumed to be arbitrarily small without violating any constraints~\cite{DEramo:2020gpr}.

 Lastly, as we observe in Fig.~\ref{fig:d5_obs}, the combination of different cosmological and astrophysical constraints lead us to consider rather large values of the reheating temperature. The current cosmic microwave background  constraint on the scalar-to-tensor ratio $(r<0.056)$ \cite{Planck:2018jri}, combined with the requirement that the inflaton decays during the inflationary period, imposes an upper bound on the reheating temperature which is roughly of order $10^{16}\GeV$. In this sense, the region of the parameter space where $T_{\text{RH}}$ approaches close to $10^{17}\GeV$ may be considered to be only marginally cosmologically acceptable.

Note that throughout the parameter space depicted in both panels of Figure \ref{fig:Model_1_res}, the DM abundance constraint can be satisfied by appropriately choosing the DM mass.

\subsection{Fermion DM}\label{sec:fermionDM}

Let us, now, turn to the case of a Dirac fermion DM candidate $\chi$, which is pair-created through a dimension-6 operator. The interaction Lagrangian that we consider is
\begin{equation}\label{eq:lagdim6}
\mathcal{L}\;\supset\;\frac{\lambda_{1}}{2\Lambda^2}\,\left(\bar{e}\PLH F_{1}\right)\,\left(\bar{\chi}\PRH\chi^c\right)\,+\,\frac{\lambda_{2}}{2\Lambda^2}\,\left(\bar{e}\PLH F_{2}\right)\,\left(\bar{\chi}\PRH\chi^c\right)\,+\,\frac{\kappa}{\Lambda^2}\,\left(\bar{e}\PLH F_{1}\right)\left(\bar{F}_2\PRH e\right)\,+\,\text{H.c.}
\end{equation}
Such a Lagrangian can, like in the case of the model described in Section \ref{sec:scalarDM}, be understood as stemming from a theory in which a scalar mediator has been integrated out. DM stability can, again, be ensured by imposing a $\mathbb{Z}_3$ symmetry under which the $F_i$'s and $\chi$ are charged according to the assignments summarized in Table \ref{tab:d6_quantum_numbers}.

\begin{table}[h]
    \centering
    \begin{tabular}{l l l}\hline\hline
         Particle & Gauge & $\mathbb{Z}_3$\\
         \hline
         $\chi$     & $(1,1)_{0}$ & $\omega$ \\
         $\bar{\chi}$     & $(1,1)_{0}$ & $\omega^{-1}$ \\
         $F_i$         & $(1,1)_{-1}$ & $\omega^{-1}$ \\
         $\bar{F}_i$   & $(1,1)_{1}$ & $\omega$ \\
        \hline\hline
    \end{tabular}
    \caption{Quantum charges of the various states, where $\omega=e^{i2\pi/3}$.}
    \label{tab:d6_quantum_numbers}
\end{table}
Much like before, $1\leftrightarrow 3$ decay processes $F_i\leftrightarrow e\chi\chi$ and $F_2\leftrightarrow F_1e\bar{e}$ are expected to be highly suppressed for $M_i \ll T_{\rm RH}$ and $\Lambda\gtrsim 10^{11}\GeV$ (see Figure \ref{fig:constraints}), with respect to $2\leftrightarrow 2$ scattering ones. The corresponding decay widths can be approximated to scale as
\begin{equation}
\Gamma_{F_i\rightarrow e\chi\chi}\,\approx\,\dfrac{\left|\lambda_{i}\right|^2}{12288 \ \pi^3}\frac{M_i^5}{\Lambda^4}
,\qquad 
\Gamma_{F_2\rightarrow F_1e\bar{e}}\,\approx\,\dfrac{\left|\kappa\right|^2}{6144 \ \pi^3}\frac{M_2^5}{\Lambda^4} 
\;,
\end{equation}
where we have assumed that $M_1\ll M_2$.

The dominant processes for the generation of the DM abundance (and the $CP$ asymmetries) are the $2\leftrightarrow 2$ scatterings $F_i\bar{e}\leftrightarrow\chi\chi$ and $F_i\bar{e}\leftrightarrow F_j\bar{e}$. An analytical approximation for the corresponding scattering rate densities can be obtained in the limit $\Lambda\gg m_i$, yielding
\begin{equation}
\gamma^{F_i\bar{e}}_{\chi\chi}\,\approx\,\frac{\left|\lambda_i\right|^2T}{512\,\pi^5\,\Lambda^4}\int_{0}^{\infty}\text{d}\tilde{s}\,\,\tilde{s}^{5/2}\,K_1\left(\frac{\sqrt{\tilde{s}}}{T}\right)\,=\,\frac{3\left|\lambda_i\right|^2}{2\,\pi^5}\frac{T^8}{\Lambda^4}
\end{equation}
whereas $\gamma^{F_i\bar{e}}_{F_j\bar{e}}$ is given by Eq.~\eqref{eq:tree-level analytical rate density model 1}.

The freeze-in DM abundance in this model can, again, be read-off from Eq.~\eqref{eq:YDMT} setting $n=2$ and $A=|\lambda_1|^2+|\lambda_2|^2$, \ie
\begin{equation}
Y_{\text{DM}}\left(T\right)\,=\,\frac{45\big(\left|\lambda_1\right|^2+\left|\lambda_2\right|^2\big)M_{Pl}}{2\times 1.66\pi^7g_{\star s}\sqrt{g_{\star\rho}}}\frac{T_{\text{RH}}^3-T^3}{\Lambda^4} \; .
\end{equation}
The DM relic density $\Omega_{\text{DM}}h^2$ can, in turn, be cast into the simple form
\begin{equation}
    \Omega_{\rm DM} h^2 \approx 0.12\times\Big(|\lambda_1|^2+|\lambda_2|^2\Big)
    \Big(\dfrac{2.3 \times 10^{15}~\GeV}{\Lambda}\Big)^4
    \Big(\dfrac{T_{\rm RH}}{3\times 10^{14}~\GeV}\Big)^3 
    \Big(\dfrac{m_{\chi}}{10 ~\keV}\Big) \;.
    \label{eq:relic_d6}
\end{equation}

\subsubsection{$CP$ asymmetries}\label{sec:CPadim6}

The $CP$ asymmetries generated by the $2\leftrightarrow 2$ scattering processes can be defined, similarly to the previous scenario that we examined, as
\begin{subequations}
\begin{alignat}{2}
\epsilon_i\,&\equiv\frac{\gamma_{F_i\bar{e}\rightarrow \chi\chi}\,-\,\gamma_{\bar{F}_i e\rightarrow\bar{\chi}\bar{\chi}}}{\gamma_{F_i\bar{e}\rightarrow \chi\chi}\,+\,\gamma_{\bar{F}_i e\rightarrow\bar{\chi}\bar{\chi}}}\label{eq:CP asymmetries 1,2 model 2}
\\\nonumber\\
\epsilon_3\,&\equiv\frac{\gamma_{F_1\bar{e}\rightarrow F_2\bar{e}}\,-\,\gamma_{\bar{F}_1 e\rightarrow\bar{F}_2 e}}{\gamma_{F_1\bar{e}\rightarrow F_2\bar{e}}\,+\,\gamma_{\bar{F}_1 e\rightarrow\bar{F}_2 e}}\label{eq:CP asymmetry 3 model 2}
\end{alignat}
\end{subequations}
where $i=1,2$. 

The imaginary part of the amplitudes can be calculated from the optical theorem to be (see Appendix \ref{sec:Appendix})

\begin{equation}
\text{Im}\big\{\mathcal{A}_0^*\mathcal{A}_1\big\}_{F_i\bar{e}\rightarrow\chi\chi}\,=\,\frac{\big(\tilde{s}-M_i^2-m_e^2\big)\big(\tilde{s}-M_j^2-m_e^2\big)\left(\tilde{s}-2m_{\chi}^2\right)\,\sqrt{\lambda\big(\tilde{s},M_j^2,m_{e}^2\big)}}{16\pi\tilde{s}} \; .
\end{equation}
Once again, this expression has been crossed-checked both using {\tt Package-X} and through the {\tt FeynRules-FeynArts-FormCalc} chain. 

\begin{figure}[t]
\centering
\includegraphics[width=0.7\linewidth]{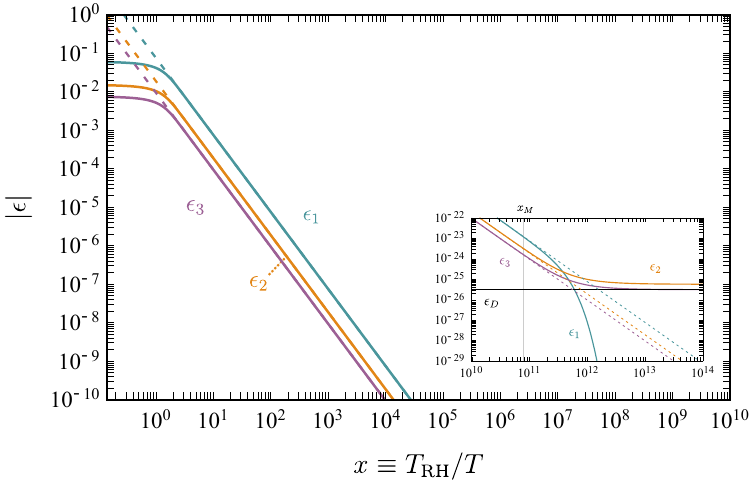}
\caption{The $CP$ asymmetries generated by the scattering processes for $\Lambda=2.85\times 10^{15}\GeV$, $T_{\text{RH}}=4\times10^{14}\GeV$, $M_1=1\TeV$, $M_2=5\TeV$, $|\lambda_1|=0.5$, $|\lambda_2|=|\kappa|=1$ and $\Delta\phi=\pi/2$. The colored solid lines correspond to the numerical results, the dashed lines to the analytical approximations, while the horizontal black solid line corresponds to the $CP$ asymmetry from the heavier fermion decay $F_2\rightarrow e\chi\chi$, \textit{i.e.} $\epsilon_D\sim\frac{\left|\lambda_2\right|^2}{32\pi}\frac{M_2^2}{\Lambda^2}$.}
\label{fig:CP Asymmetries model 2}
\end{figure}

An analytical approximation for the $CP$ asymmetries $\epsilon_1$ and $\epsilon_2$ can be obtained in the limit $\Lambda\gg m_i$, yielding
\begin{subequations}
\begin{align}
\epsilon_1\,&\approx\,-\frac{6}{\pi}\frac{\left|\kappa\right|\left|\lambda_2\right|}{\left|\lambda_1\right|}\sin\left(-\Delta \phi \right)\frac{T^2}{\Lambda^2}
\\\nonumber\\
\epsilon_2\,&\approx\,-\frac{6}{\pi}\frac{\left|\kappa\right|\left|\lambda_1\right|}{\left|\lambda_2\right|}\sin\left(\Delta \phi\right)\frac{T^2}{\Lambda^2} \; .
\end{align}
\end{subequations}
In the case of the $\epsilon_3$ asymmetry the imaginary part of the Feynman amplitudes is

\begin{equation}
\text{Im}\big\{\mathcal{A}_0^*\mathcal{A}_1\big\}_{F_1\bar{e}\rightarrow F_2\bar{e}}\,=\,\frac{\big(\tilde{s}-M_1^2-m_e^2\big)\big(\tilde{s}-M_2^2-m_e^2\big)\left(\tilde{s}-2m_{\chi}^2\right)\,\sqrt{\lambda\big(\tilde{s},m_{\chi}^2,m_{\chi}^2\big)}}{16\pi\tilde{s}}
\end{equation}
and can be analytically approximated by
\begin{equation}
\epsilon_3\,\approx\,-\frac{3}{\pi}\frac{\left|\lambda_1\right|\left|\lambda_2\right|}{\left|\kappa\right|}\sin\left(\Delta \phi\right)\frac{T^2}{\Lambda^2} \; .
\end{equation}

In Figure \ref{fig:CP Asymmetries model 2} we illustrate the values of the $CP$ asymmetries as a function of $x \equiv T_{\rm RH}/T$, for the representative set of parameter values: $\Lambda=2.85\times10^{15}\GeV$, $T_{\text{RH}}=4\times10^{14}\GeV$, $M_1=1\TeV$, $M_2=5\TeV$, $|\lambda_1|=0.5$, $|\lambda_2|=|\kappa|=1$ and $\Delta\phi=\pi/2$. Again, the numerical and analytical results are identical at the temperatures of interest. 

Finally, invariance under $CPT$ and $S$-matrix unitarity result to the same conditions as before, Eq.~\eqref{eq:CPT+unitarity conditions model 1}, with $\varphi$ replaced by the Dirac DM state $\chi$.

\subsubsection{Baryon asymmetry}

The Boltzmann equations describing the temperature evolution of the asymmetries in the $F_i$'s and in the SM can be written in a completely analogous manner as in the scalar DM case. We repeat them here for convenience:
\begin{subequations}\label{eq:d6-Boltzmann equations}
\begin{align}
-sHT\frac{\text{d}Y_{\Delta F_i}}{\text{d}T}\,&=\,\epsilon_i\gamma^{F_i\bar{e}}_{\chi\chi}\,-\,\frac{1}{2}\gamma^{F_i\bar{e}}_{\chi\chi}\left(y_{F_i}-y_e\right)\,+(-1)^i\,\gamma^{F_1\bar{e}}_{F_2\bar{e}}\left(y_{F_1}-y_{F_2}\right)
\label{eq:d6-BEYFi}\\\nonumber\\
-sHT\frac{\text{d}Y_{B-L_{\text{SM}}}}{\text{d}T}\,&=\,-\frac{1}{2}\gamma^{F_1\bar{e}}_{\chi\chi}\left(y_{F_1}-y_e\right)\,-\,\frac{1}{2}\gamma^{F_2\bar{e}}_{\chi\chi}\left(y_{F_2}-y_e\right) \, . 
\label{eq:d6-BEYDFi}
\end{align}
\end{subequations}

\begin{figure}[t]
\centering
\includegraphics[width=0.7\linewidth]{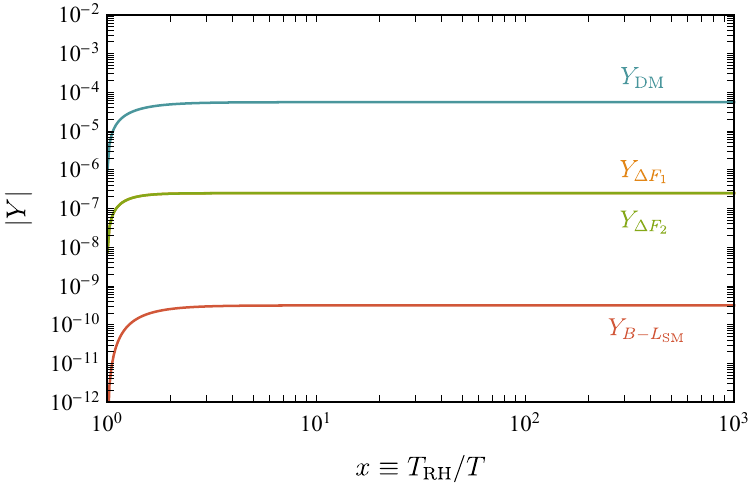}
\caption{The generated asymmetries $Y_{\Delta F_i}, Y_{B-L_{\text{SM}}}$ and the DM abundance $Y_{\text{DM}}$, in terms of the dimensionless parameter $x\equiv T_{\text{RH}}/T$.}
\label{fig:Asymmetric yields model 2}
\end{figure}
In Figure \ref{fig:Asymmetric yields model 2} we present an explicit example of the temperature evolution of the various yields, in which we have assumed the same set of parameters as those used in Figure \ref{fig:CP Asymmetries model 2}. In order to obtain the observed DM relic density, the mass of the DM state, as can be seen from Eq.~\eqref{eq:relic_d6}, is $m_{\chi}\approx 8\keV$.

\subsubsection{Results}\label{sec:resultsdim6}

\begin{figure}[t]
    \hspace*{-1.9cm}
\begin{subfigure}{.6\linewidth}
  \includegraphics[width=\linewidth]{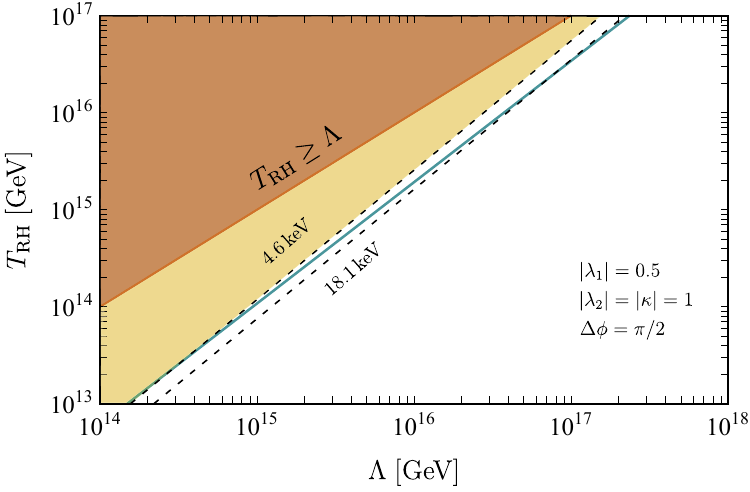}
  \caption{}
  \label{fig:Baryon asymmetry scan model 2}
\end{subfigure}\hspace*{-0.0cm}
\begin{subfigure}{.61\linewidth}
\vspace*{-0.2cm}
  \includegraphics[width=\linewidth]{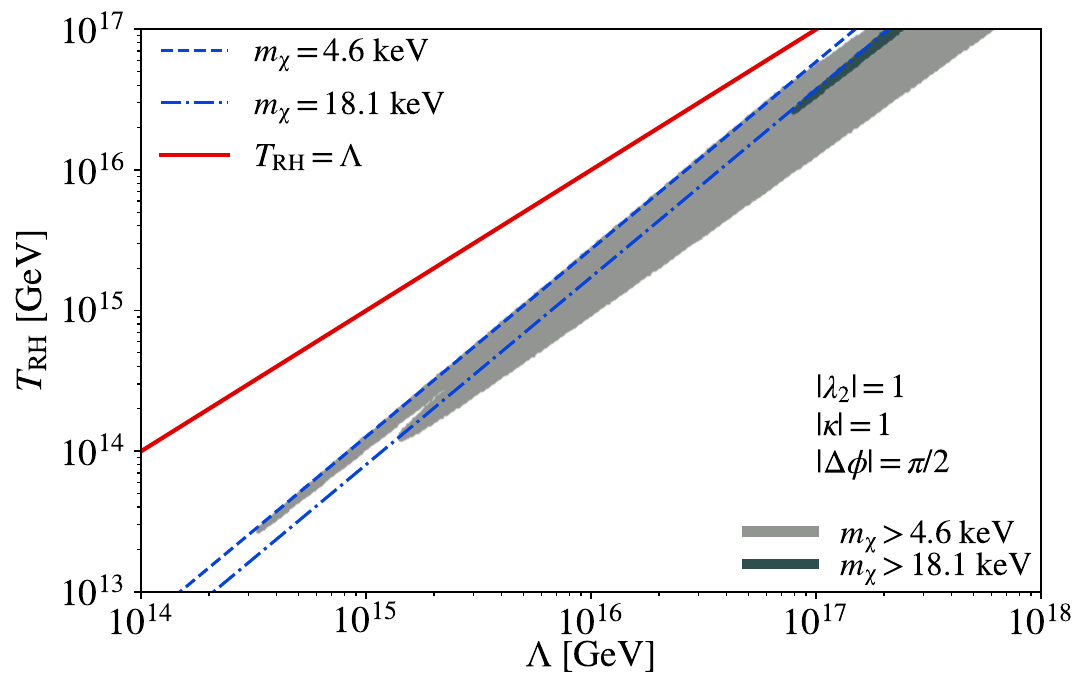}
  \caption{}
  \label{fig:d6_obs}
\end{subfigure}
\caption{(a) Combinations of the reheating temperature $T_{\text{RH}}$ and the energy scale $\Lambda$, which can generate the observed baryon asymmetry. The coupling constants have been set to $|\lambda_1|=0.5$, $|\lambda_2|=|\kappa|=1$ and $\Delta\phi=\pi/2$. The dashed lines depict representative DM masses $m_{\chi}=\{4.6,18.1\}\keV$ for which the observed DM relic density can be reproduced. Masses below $4.6\keV$ are excluded from current Lyman-$\alpha$ forest observations (yellow-shaded area). (b) The values of  $\Lambda$ and $T_{\rm RH}$ that result in the observed DM relic and baryon asymmetry for $|\kappa| = |\lambda_{1}| = 1$ and $\Delta\phi=\pi/2$, while $|\lambda_2|$ is chosen so that $\Omega_{\text{DM}} h^2 =0.12$. The darker-shaded region corresponds to corresponds to $m_{\chi} > 18.1~\keV$, and it is extended to the lighter-shaded one for $m_{\chi} > 4.6~\keV$. The blue lines show the upper bound on $T_{\rm RH}$ for which the observed DM abundance in the Universe can be reproduced for each value of $\Lambda$, for the two DM mass limits.}
\label{fig:Model_2_res}
\end{figure}

Our numerical results are summarized in Figure \ref{fig:Model_2_res} following a methodology similar to the one described in Section \ref{sec:resultsdim5}. In the left-hand side panel, we show the combinations of the reheating temperature and the EFT energy scale for which the observed baryon asymmetry can be obtained, keeping the couplings fixed at the values $|\lambda_1|=0.5$, $|\lambda_2|=|\kappa|=1$ and $\Delta\phi=\pi/2$. Here, the Boltzmann equations Eqs.~\eqref{eq:d6-Boltzmann equations} have been solved numerically, using the analytical approximations for the different rate densities and $CP$ asymmetries that we extracted in Sections \ref{sec:fermionDM} and \ref{sec:CPadim6}.

In the right-hand side panel, we again present a more extended scan of the available parameter space. We can derive an approximate solution for $Y_{B-L_{\text{SM}}}$ by keeping only the first terms of Eqs.~\eqref{eq:d6-BEYFi}. The leading-order approximation in this case reads
\begin{align}\label{eq:d6-YDe_approx}
    \frac{22}{79}Y_{B-L_{\text{SM}}} \approx  -  \, 8.71 \times 10^{-11} \, 
    |\kappa|  |\lambda_1| |\lambda_2| \lrb{|\lambda_1|^2-|\lambda_2|^2}
    \sin\lrb{\Delta \phi} \times & \\ \nonumber 
    \times \lrb{\dfrac{T_{\rm RH}}{1.45 \times 10^{15}~\GeV}}^{8}
    \lrb{\dfrac{10^{16}~\GeV}{\Lambda}}^{10} & .
\end{align}
By requiring the DM abundance and the baryon asymmetry to correspond to their respective observed values, from Eqs.~\eqref{eq:relic_d6} and \eqref{eq:d6-YDe_approx} we find that the cut-off $\Lambda$ and the reheating temperature $T_{\rm RH}$ must be related approximately through
\begin{subequations}
\begin{align}
    & \Lambda \approx 7.8 \times 10^{14}~\GeV \ 
    \dfrac{\lrb{|\lambda_1|^2+|\lambda_2|^2}^4}{\lrb{\left||\kappa||\lambda_1||\lambda_2|^2(|\lambda_1|^2-|\lambda_2|^2)\sin\lrb{\Delta \phi}  \right|}^{3/2}} 
    \lrb{\dfrac{m_{\chi}}{10~\keV}}^4
    \\\nonumber\\
    &T_{\rm RH} \approx 7 \times 10^{13} ~\GeV \ 
    \dfrac{\lrb{|\lambda_1|^2+|\lambda_2|^2}^5}{\left||\kappa||\lambda_1||\lambda_2|(|\lambda_1|^2-|\lambda_2|^2)\sin\lrb{\Delta \phi}  \right|^2} 
    \lrb{\dfrac{m_{\chi}}{10~\keV}}^5 \;.
    \label{eq:d6_obs}
\end{align}
\end{subequations}
Our results (which we have verified to agree within a $10\%$ accuracy with a fully numerical resolution of the Boltzmann equations) are shown in Fig.~\ref{fig:d6_obs}.

We observe that in this case, it is possible to satisfy the baryon asymmetry and DM abundance bounds for a fairly wider range of $\Lambda$ and $T_{\rm RH}$ with respect to the scalar DM case that we examined previously. Besides, similar remarks as the ones we made in Section \ref{sec:resultsdim5} apply in this case as well; we indeed find that the heavy fermions decay long after sphaleron decoupling which, in turn, ensures that the baryon asymmetry survives until the present day. In order to prevent the $F_i$'s from decaying too late and, hence, interfering with primordial nucleosynthesis, they must either be heavy (in this case, heavier than several thousands of TeV) or, alternatively, an additional decay channel can be introduced which enables the $F_i$'s to decay before BBN. All in all, this setup appears to offer more possibilities than the scalar DM one, since it allows for greater freedom in the choice of the EFT scale and the reheating temperature.

\section{Conclusions and Outlook}\label{sec:conclusions}

In this paper we presented a mechanism in order to explain simultaneously the observed DM abundance and matter-antimatter asymmetry of the Universe. A symmetric DM density is created through the freeze-in mechanism,  relying on highly suppressed $2 \leftrightarrow 2$ scattering processes described by non-renormalizable operators. At the same time, these scattering processes also violate $CP$ which amounts, in synergy with the action of electroweak sphalerons, to an asymmetry between SM baryons and antibaryons. The fact that the relevant interactions are non-renormalizable leads to both abundances being generated at high temperatures, conversely to other scenarios of freeze-in baryogenesis which have been considered in the literature \cite{Shuve_2020, Goudelis:2021qla, Chand:2022vrf}.

As a proof-of-concept we studied two simple scenarios of scalar and fermion DM, in which the interactions between the  dark and visible particles are described by dimension-5 and dimension-6 operators, respectively, involving the DM particles themselves, SM fermions and exotic vector-like fermions. $CP$ violation is generated through interference between the leading-order and next-to-leading-order Feynman diagrams. Our numerical analysis showed that in both cases it is, indeed, possible to simultaneously freeze-in the necessary DM density along with a viable matter-antimatter asymmetry, as long as DM is relatively light (but within existing Lyman-$\alpha$ forest bounds), the interaction scale $\Lambda$ is larger than $O(10^{16})$ [$O(10^{14})$] GeV and the reheating temperature $T_{\rm RH}$ is higher than $O(10^{15})$ [$O(10^{13})$] GeV in the case of scalar (fermion) DM. The masses of the vector-like fermions can be quite low, down to a few TeV, although in the case of fermion DM this would require the existence of an additional decay channel so that they decay before the era of BBN. 

Interestingly, in such a scenario (\textit{i.e.} of relatively light vector-like fermions) there could be favorable prospects for the detection of these particles at the Large Hadron Collider. Concretely, the heavy fermions in the two scenarios that we studied, if kinematically accessible, should be copiously Drell-Yan-produced at the LHC due to their gauge interactions. Then, depending on the precise value of the lifetime, these particles can subsequently either decay -- typically displaced with respect to the primary interaction point -- into SM leptons accompanied by missing energy or manifest themselves as Heavy (meta-)Stable Charged Particles (HSCPs), with the latter being the favored case in the models that we studied (for an overview of the physics opportunities for LLP searches at the LHC \textit{cf} \cite{Alimena:2019zri}). According to the analysis performed, \textit{e.g.}, in \cite{Belanger:2018sti}, such scenarios can be probed at the high-luminosity run of the LHC for heavy fermion masses up to $\sim 1.5$ TeV (or heavier if they also carry color). Although we by no means claim that the LHC can fully scrutinize our proposal, it is remarkable that at least part of the parameter space can give rise to such observable phenomenological signatures. On the side of cosmology, on the other hand, the two scenarios that we studied can simultaneously explain dark matter and the baryon asymmetry of the Universe as long as dark matter is relatively light. This feature could have interesting implications for astrophysical observations, most notably the small-scale structure issues of $\Lambda$CDM.

There are several ways through which our analysis could be extended. First of all, the simple scenarios that we presented were intended to serve mostly as proofs-of-concept concerning the fact that successful freeze-in baryogenesis can be realized in the UV. Clearly, much more elaborate models can be developed, based on well-motivated extensions of the SM. In a similar vein, it would be interesting to examine whether the parameter space can be extended to lower cutoff and/or reheating temperature values as well as to accommodate heavier DM. Indeed, preference for relatively light DM appears to be a common (albeit to different extents, depending on the precise mechanism that is responsible for $CP$ violation) in the freeze-in baryogenesis scenarios that have appeared in the literature so far \cite{Shuve_2020, Goudelis:2021qla}. Is this a generic feature, or not? We hope to address these questions in future work.

\acknowledgments 
The authors acknowledge useful discussions with Ioannis Dalianis. 
D.K. acknowledges support by the Lancaster-Manchester-Sheffield Consortium for Fundamental Physics under STFC Grant No. ST/T001038/1. The  work of V.C.S. was supported by the Hellenic Foundation for Research and Innovation (H.F.R.I.) under the ``First Call for H.F.R.I. Research Projects to support Faculty members and Researchers and the procurement of high-cost research equipment grant'' (Project Number: 824).  This research is co-financed by Greece and the European Union (European Social Fund - ESF) through the  Operational Programme ``Human Resources Development, Education and Lifelong Learning'' in the context of the project ``Strengthening Human Resources Research Potential via Doctorate Research - 2nd Cycle'' (MIS-5000432), implemented by the State Scholarships Foundation (IKY).

\appendix

\section{Loop Calculations with Optical Theorem}\label{sec:Appendix}

In our model $CP$-violating effects arise, at lowest order, due to the interference between tree-level and 1-loop diagrams. In addition, the intermediate states in the loop must propagate on-shell so that the corresponding Feynman amplitudes obtain an imaginary part \cite{Kolb:1979qa}. A direct way to calculate them is by using the optical theorem. In this appendix we present more details on the calculations of the imaginary parts of the $\epsilon_i (i=1,2)$ and $\epsilon_3$ $CP$ asymmetries that appear in the models under consideration.
\\
\begin{enumerate}[i.]
    \item $F_i\bar{e}\rightarrow F_j\bar{e}\rightarrow \varphi\varphi$.
\suspend{enumerate}
\begin{align}
2i\text{Im}\big\{\mathcal{A}_0^*\mathcal{A}_1\big\}&=i\mathcal{A}_0^*\left(F_i\bar{e}\rightarrow\varphi\varphi\right)\int\text{d}\Pi_{F_j}\text{d}\Pi_{\bar{e}}\left(2\pi\right)^4\delta^{(4)}\mathcal{A}_0\left(F_i\bar{e}\rightarrow F_j\bar{e}\right)\mathcal{A}_0\left(F_j\bar{e}\rightarrow\varphi\varphi\right)\nonumber
\\
&=4i\int\text{d}\Pi_{F_j} \, \text{d}\Pi_{\bar{e}} \, \left(2\pi\right)^4\delta^{(4)}\left(p_{F_i}\cdot p_{\bar{e}}\right)\left(p_{F_j}\cdot p_{\bar{e}}\right)\nonumber
\\
&=\frac{i}{8\pi\tilde{s}}\big(\tilde{s}-M_i^2-m_e^2\big)\big(\tilde{s}-M_j^2-m_e^2\big)\,\sqrt{\lambda\big(\tilde{s},M_j^2,m_{e}^2\big)}
\end{align}
\\

\resume{enumerate}[{[i.]}]
    \item $F_i\bar{e}\rightarrow \varphi\varphi\rightarrow F_j\bar{e}$.
\suspend{enumerate}
\begin{align}
2i\text{Im}\big\{\mathcal{A}_0^*\mathcal{A}_1\big\}&=i\mathcal{A}_0^*\left(F_i\bar{e}\rightarrow F_j\bar{e}\right)\int\text{d}\Pi_{\varphi} \, \text{d}\Pi_{\varphi}\, \left(2\pi\right)^4\delta^{(4)}\mathcal{A}_0\left(F_i\bar{e}\rightarrow\varphi\varphi\right)\mathcal{A}_0\left(\varphi\varphi\rightarrow F_j\bar{e}\right)\nonumber
\\
&=4i\int\text{d}\Pi_{\varphi} \, \text{d}\Pi_{\varphi}\, \left(2\pi\right)^4\delta^{(4)}\left(p_{F_i}\cdot p_{\bar{e}}\right)\left(p_{F_j}\cdot p_{\bar{e}}\right)\nonumber
\\
&=\frac{i}{8\pi\tilde{s}}\big(\tilde{s}-M_i^2-m_e^2\big)\big(\tilde{s}-M_j^2-m_e^2\big)\,\sqrt{\lambda\big(\tilde{s},m_{\varphi}^2,m_{\varphi}^2\big)}
\end{align}
\\

\resume{enumerate}[{[i.]}]
    \item $F_i\bar{e}\rightarrow F_j\bar{e}\rightarrow \chi\chi$.
\suspend{enumerate}
\begin{align}
2i\text{Im}\big\{\mathcal{A}_0^*\mathcal{A}_1\big\}&=i\mathcal{A}_0^*\left(F_i\bar{e}\rightarrow\chi\chi\right)\int\text{d}\Pi_{F_j}\, \text{d}\Pi_{\bar{e}} \, \left(2\pi\right)^4\delta^{(4)}\mathcal{A}_0\left(F_i\bar{e}\rightarrow F_j\bar{e}\right)\mathcal{A}_0\left(F_j\bar{e}\rightarrow\chi\chi\right)\nonumber
\\
&=8i\int\text{d}\Pi_{F_j}\, \text{d}\Pi_{\bar{e}} \, \left(2\pi\right)^4\delta^{(4)}\left(p_{F_i}\cdot p_{\bar{e}}\right)\left(p_{\chi_1}\cdot p_{\chi_2}\right)\left(p_{F_j}\cdot p_{\bar{e}}\right)\nonumber
\\
&=\frac{i}{8\pi\tilde{s}}\big(\tilde{s}-M_i^2-m_e^2\big)\big(\tilde{s}-M_j^2-m_e^2\big)\left(\tilde{s}-2m_{\chi}^2\right)\,\sqrt{\lambda\big(\tilde{s},M_j^2,m_{e}^2\big)}
\end{align}
\\

\resume{enumerate}[{[i.]}]
    \item $F_i\bar{e}\rightarrow \chi\chi\rightarrow F_j\bar{e}$.
\end{enumerate}
\begin{align}
2i\text{Im}\big\{\mathcal{A}_0^*\mathcal{A}_1\big\}&=i\mathcal{A}_0^*\left(F_i\bar{e}\rightarrow F_j\bar{e}\right)\int\text{d}\Pi_{\chi} \, \text{d}\Pi_{\chi} \, \left(2\pi\right)^4\delta^{(4)}\mathcal{A}_0\left(F_i\bar{e}\rightarrow\chi\chi\right)\mathcal{A}_0\left(\chi\chi\rightarrow F_j\bar{e}\right)\nonumber
\\
&=8i\int\text{d}\Pi_{\chi}\, \text{d}\Pi_{\chi}\, \left(2\pi\right)^4\delta^{(4)}\left(p_{F_i}\cdot p_{\bar{e}}\right)\left(p_{F_j}\cdot p_{\bar{e}}\right)\left(p_{\chi_1}\cdot p_{\chi_2}\right)\nonumber
\\
&=\frac{i}{8\pi\tilde{s}}\big(\tilde{s}-M_i^2-m_e^2\big)\big(\tilde{s}-M_j^2-m_e^2\big)\left(\tilde{s}-2m_{\chi}^2\right)\,\sqrt{\lambda\big(\tilde{s},m_{\chi}^2,m_{\chi}^2\big)}
\end{align}


\bibliographystyle{JHEP}
\bibliography{bibliography_UVFIBG}

\end{document}